# Quantum advantage of nonlinear quantum battery and superconducting circuit implementation

**Wei-Jun Han, Peng-Yu Sun, Guo-Feng Zhang***

*School of Physics, Beihang University, Beijing 100191, China*

**Abstract:** A quantum battery is a novel energy storage device that operates on the principles of quantum mechanics. To enhance the charging performance of quantum batteries and further provide theoretical support for their physical implementation, we constructed an optical-field-dependent nonlinear quantum battery model. Meanwhile, we solved for the unbiased form of nonlinear interactions in this model, where the charging power of the proposed model exhibits a superlinear quantum advantage, and the charging time saturates the quantum speed limit. Through theoretical analysis, we confirm that this quantum advantage arises from the quantum effect of multiphoton absorption. Subsequently, with the derived nonlinear function form, we further investigated other properties of this nonlinear quantum battery. Finally, an experimental design scheme for this nonlinear quantum battery in superconducting quantum circuits is presented.

## 1 Introduction

Quantum thermodynamics is a subfield focused on investigating thermodynamic processes, including energy conversion, storage, and work extraction, in quantum systems. A quantum battery (QB) is a microscopic quantum device that can temporarily store and release energy in a controllable manner, thus serving as an ideal platform for advancing quantum thermodynamics research. Research on QBs primarily follows two key directions. First, it explores correlations between QB performance and various quantum resources, with researchers leveraging these resources to enhance charging efficiency [1-9]. The second direction involves proposing diverse QB models, such as spin chain systems [4,10], the Tavis-Cummings model [11,12], harmonic oscillators [13], the Dicke model [14,15], and three-level systems [16-18]. Concurrently, numerous charging protocols have been developed, including parallel and collective charging [14,19], wireless charging [20], cavity-assisted charging [21,22], and environment-assisted charging [23-26]. Among these, collective charging in Dicke quantum batteries exhibits a superlinear power advantage; however, this advantage is shown to stem from classical collective effects rather than intrinsic quantum phenomena. Models with genuine quantum advantage remain rare. Currently, only the Sachdev-Ye-Kitaev (SYK) model [19] and the anharmonic bosonic quantum battery model [27] have been confirmed to possess such advantage. In addition, the quantum advantage of Ref.27 originates from the intrinsic anharmonicity of the system itself. It is one of the pioneering works that introduce nonlinear components into QB. This work [27] establishes a fair comparison benchmark between linear and nonlinear QBs, and provides significant methodological guidance for the subsequent research on QB with nonlinear effects.

We define a QB that enhances charging performance by introducing nonlinear interactions between the charger and the battery as a nonlinear QB. Then, two approaches exist for incorporating nonlinearity into this coupling: The first, exemplified by works [28-30], embeds nonlinear functions into the linear coupling strength relationship. However, these studies artificially predefine the form of nonlinear interactions, which may not correspond to the optimal configuration. The second approach focuses on energy exchange mechanisms, such as two-photon quantum batteries [31-33] and nonlinear two-photon coupling analogous to degenerate parametric down-conversion [34]. Building on these prior advancements, this paper presents a comprehensive study of nonlinear quantum battery: We integrate the two aforementioned approaches for introducing nonlinear interactions, which encompass both energy transfer regulated by nonlinear functions and multiphoton absorption as an energy exchange mechanism. Accordingly, in this work, we construct an optical-field-dependent nonlinear quantum battery (O-NQB). In our model, we do not predefine the specific form of the optical-field-dependent nonlinear function $f(n)$. Inspired by Ref.27, we adopt the same comparison benchmark. Then, we derive an unbiased $f(n)$, which avoids spurious charging advantages caused by arbitrary manual parameter tuning. Finally, we find that the charging time of the O-NQB can equal the quantum speed limit, and that the battery's charging power exhibits a superlinear quantum advantage. Since this advantage arises under the unbiased form of $f(n)$, the quantum advantage in charging stems from the introduction of multiphoton absorption, which is a quintessential nonlinear quantum effect, and is fundamentally different in physical origin from the quantum advantage in Ref. [27], which originates from the intrinsic anharmonicity of bosonic oscillators.

*Contact author: gf1978zhang@buaa.edu.cn

Next, we discuss the current state of QB research from the perspective of physical implementation. Experimental QB research is currently in the phase of laboratory prototype validation and key technology breakthroughs. Promising physical platforms for QB realization include superconducting quantum circuits [35-38], quantum dots [39-41], organic microcavities [42], and nuclear spins [43]. Recently, a theoretical scheme centered on diamond nitrogen-vacancy (NV) centers was proposed to mitigate QB self-discharge and optimize extractable work, providing critical support for practical NV-center-based QBs [44]. In this work, we use superconducting quantum circuits to design an experimental scheme for our proposed O-NQB. This choice is motivated by three factors: First, superconducting quantum circuit technology is relatively mature. Second, in Ref. [19] envisions QBs as potential power sources for quantum computers, which is a long-standing goal for researchers dedicated to exploring their practical applications. Notably, superconducting quantum circuits themselves are also one of the leading platforms for quantum computing, and this aligns our QB design with this future application. Finally, our model centers on multi-photon absorption regulated by optical-field intensity, and Josephson junctions (key components of superconducting circuits) possess intrinsic high-order nonlinearity [35]. Furthermore, our proposed scheme is structurally simple, facilitating laboratory implementation, and may accelerate progress toward QB-powered quantum computers, ultimately providing a feasible experimental pathway to unlock practical QB applications.

The structure of this paper is as follows: In Sect. 2, details the O-NQB. In Sect. 3, derives the unprejudiced form of the nonlinear interaction function in this model and verifies the O-NQB's quantum advantage. In Sect.4, discuss the physical mechanism of the quantum advantage. In Sect. 5, analyzes charging performance metrics of this O-NQB under the objective nonlinear function, as well as the impact of detuning on charging. And in Sect. 6, presents the experimental design for our model based on superconducting quantum circuits. Finally, we present summaries and conclusions in Sect. 7.

## 2 Model

This section details the proposed O-NQB, where the optical cavity acts as the charger and the two-level atom inside it serves as the battery. The interaction between the charger and the battery is not a conventional linear interaction but a nonlinear one dependent on the optical field intensity in the cavity. Its Hamiltonian is given by ($\hbar = 1$ throughout this paper)

$$H = \omega a^{\dagger} a + \frac{1}{2}\Omega\sigma_z + \theta(t) g\left[A^k \sigma^+ + \left(A^{\dagger}\right)^k \sigma^-\right] \tag{1}$$

Here, $a^{\dagger}(a)$ denotes the creation (annihilation) operator, $\sigma^+(\sigma^-)$ denotes the raising (lowering) operator of the two-level atom, and $\sigma_z = \sigma^+\sigma^- - \sigma^-\sigma^+$. Additionally, $\omega$ is the frequency of the intracavity optics field, $\Omega$ is the transition frequency of the two-level atom within the cavity, and $g$ is the coupling constant between the cavity and the atom. Finally, $A = af(a^{\dagger}a)$ and $A^{\dagger} = f(a^{\dagger}a)a^{\dagger}$, where $f(a^{\dagger}a)$ is defined as a nonlinear interaction function dependent on the optical field intensity. The third term in Eq. (1) represents the nonlinear interaction, whose physical meaning is that the energy of multiple photons in the charger is simultaneously transferred to the battery, inducing an energy level transition in the battery. However, since the nonlinear function acts continuously during the transition, the form $[af(a^{\dagger}a)]^k$ is adopted instead of $a^k f(a^{\dagger}a)$, which is also more mathematically rigorous. Additionally, in Eq. (1), when $k = 1$ and $f(a^{\dagger}a) = 1$, this model reduces to the standard Jaynes-Cummings linear quantum battery model (JC-LQB) where no nonlinear effects operate, i.e., $H \to H_{\mathrm{L}}$. By contrast, when $k \geq 2$ and $f(a^{\dagger}a) \neq 1$, nonlinear effects come into play, which corresponds to our proposed O-NQB model. The specific form of $f(a^{\dagger}a)$ will be detailed in the next section.

The schematic diagram of the specific model is shown in Fig. 1, where $\theta(t)$ is the control switch of the charger, which takes a value of 1 during the charging process in the interval $(0, t_f]$ and 0 after the battery is fully charged.

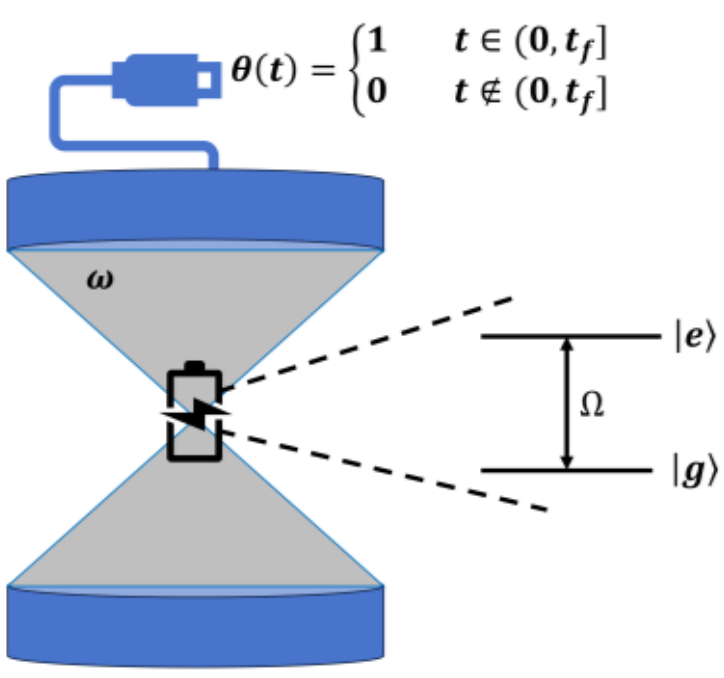

FIG.1. Schematic of the optical-field-dependent nonlinear quantum battery model, where the optical cavity acts as the charger and the two-level atom inside the cavity serve as the battery.

For the next purpose, it is useful to express the Hamiltonian in the interaction picture, which is given by

$$V_I(t) = \theta(t) g\left[A^k \sigma^+ e^{-i\Delta t} + \left(A^{\dagger}\right)^k \sigma^- e^{i\Delta t}\right] \tag{2}$$

with $\Delta = k\omega - \Omega$.

We assume the initial state of this O-NQB is $|\psi_I(0)\rangle = |n, g\rangle$, where the optical field contains

$n$ photons and the two-level atom is in the ground state. The time-evolved wave function of the system in the interaction picture is then $|\psi_I(t)\rangle = c_1(t)|n,g\rangle + c_2(t)|0,e\rangle$, with $c_1(0) = 1$ and $c_2(0) = 0$. The wave function $|\psi_I(t)\rangle$ in the interaction picture is determined by solving the Schrodinger equation in the interaction picture, expressed as

$$i\frac{\partial}{\partial t}|\psi_I(t)\rangle = V_I(t)|\psi_I(t)\rangle \tag{3}$$

$|\psi_I(t)\rangle$ is substituted into Eq. (3) yields

$$\begin{aligned} i\dot{c}_1(t) &= ge^{i\Delta t}\xi c_2(t) \\ i\dot{c}_2(t) &= ge^{-i\Delta t}\xi c_1(t) \end{aligned} \tag{4}$$

Where, $\xi = \left(\prod_{j=1}^{k}\sqrt{j}\,f(j)\right)$. Further solving Eq. (4) yields

$$\begin{aligned} c_1(t) &= e^{i\frac{\Delta t}{2}}\left[\cos\left(\frac{R_n(\Delta)}{2}t\right) - i\frac{\Delta}{R_n(\Delta)}\sin\left(\frac{R_n(\Delta)}{2}t\right)\right] \\ c_2(t) &= -ie^{-i\frac{\Delta t}{2}}\frac{R_n(0)}{R_n(\Delta)}\sin\left(\frac{R_n(\Delta)}{2}t\right) \end{aligned} \tag{5}$$

Here, $R_n(\Delta) = \sqrt{4g^2\xi^2 + \Delta^2}$.

Thus, the density matrix of the O-NQB is obtained by tracing over the field degrees of freedom $\rho_{\mathrm{B}} = \mathrm{Tr}_{\mathrm{field}}(|\psi_I(t)\rangle\langle\psi_I(t)|)$, given by

$$\rho_{\mathrm{B}}(t) = |c_1(t)|^2|g\rangle\langle g| + |c_2(t)|^2|e\rangle\langle e| \tag{6}$$

**3 Optical-field-dependent nonlinear interaction function**

Since we have not predefined the form of the nonlinear function, our core task is to derive its form. So, this section focuses on discussing the specific form of the nonlinear interaction function.

First, we introduce key metrics for QB charging performance: stored energy $\Delta E(t)$, charging time $t_f$, and average charging power $\langle P(t)\rangle$. The $\Delta E(t)$ of the QB is defined as the difference in the average energy of the battery unit before and after charging

$$\Delta E(t) = \mathrm{Tr}[\rho_{\mathrm{B}}(t)H_{\mathrm{B}}] - \mathrm{Tr}[\rho_{\mathrm{B}}(0)H_{\mathrm{B}}] \tag{7}$$

here, $H_{\mathrm{B}} = \frac{1}{2}\Omega\sigma_z$ is the Hamiltonian of the battery unit. We terminate charging as soon as $\Delta E(t)$ reaches its first peak $\Delta E_{\max} = \Delta E(t_f)$; the corresponding time $t_f$ is called the charging time. From this, we readily define the average power as $\langle P(t)\rangle = \frac{\Delta E(t)}{t}$, so the average power over the entire charging process is

$$\langle P(t_f)\rangle = \frac{\Delta E_{\max}}{t_f} \tag{8}$$

Next, we analyze the impact of nonlinear interactions on QB performance. To better illustrate the role of the nonlinear function, we compare the charging performance of JC-LQB and O-NQB to assess whether nonlinear interactions can enhance the charging performance. However, directly comparing the charging times $t_f$ of the two models would be unfair, as arbitrary choices of $f(n)$ (e.g., setting $f(n) \gg g$) could shorten the nonlinear battery's charging time via artificial parameter tuning (not quantum effects). Therefore, establishing a fair comparison benchmark is crucial. Under this benchmark, the derived form of $f(n)$ can avoid unfairness caused by arbitrary parameter settings.

In this work, drawing on the method of Ref.27, we select the alignment of the quantum speed limit (QSL) [19,45-48] as the core benchmark for model comparison, with the full definition and the rationale for this selection elaborated as follows: The QSL, as dictated by the fundamental principles of quantum mechanics, defines the theoretical lower bound on the minimum time required for a quantum system to evolve from its initial state to an orthogonal final state. It is an intrinsic property of quantum systems, entirely independent of the coupling strength, the form of the nonlinear function, and any artificially tunable parameters. Aligning the QSL of the linear and nonlinear models enables us to completely eliminate spurious performance advantages introduced by artificial parameter tuning, and leaves the nonlinear quantum effect of multiphoton absorption as the sole variable between the two model frameworks. This allows for a rigorous verification of the intrinsic performance improvement induced by this effect, establishing QSL alignment as the only unbiased benchmark capable of isolating the intrinsic contribution of the quantum effect. Accordingly, we assume that the theoretical lower bounds of the charging times for the two models are equal, i.e., we equate the QSL times of the two models, to establish a unified and fair comparison benchmark. Finally, we demonstrate the uniqueness and rigor of selecting QSL alignment as the comparison benchmark from its specific mathematical formulation. For the O-NQB, the stored energy is given by $\Delta E(t) = \Omega|c_2(t)|^2$, which can be reduced to $\Delta E(t) = n\omega \sin^2\left[gt\left(\prod_{j=1}^{n}\sqrt{j}f(j)\right)\right]$ under the resonant case. From this expression, $f(n)$ appears in the oscillatory term and mainly functions to regulate the charging time. Therefore, the only feasible approach to eliminate the influence of $f(n)$ is to adopt a fixed charging time as the comparison benchmark. This renders QSL alignment the only unbiased benchmark that can completely eliminate the interference caused by artificial parameter tuning and accurately isolate the intrinsic quantum effect of multiphoton absorption, which constitutes the core rationale for our selection of this benchmark in this work. By contrast, adopting benchmarks of fixed coupling strength or fixed upper limit of energy storage cannot achieve this goal and is thus not applicable for the fair comparison in our study.

We first present the method for solving the QSL of a closed system

$$\tau_{\text{QSL}} = \max\left\{\frac{\pi}{2\langle\delta H\rangle_{\psi(0)}}, \frac{\pi}{2\langle H - E_0\rangle_{\psi(0)}}\right\} \tag{9}$$

Here, $\langle\delta H\rangle_{\psi(0)} = \sqrt{\langle\psi(0)|H^2|\psi(0)\rangle - \langle\psi(0)|H|\psi(0)\rangle^2}$ denotes the variance of the total Hamiltonian in the initial state. Additionally, $\langle H - E_0\rangle_{\psi(0)}$ is the difference between the system's initial average energy and its ground-state energy $E_0$. For $g < \omega\sqrt{n}$, we adopt the Mandelstam-Tamm QSL

$$\tau_{\text{QSL}} \equiv \frac{\pi}{2\langle\delta H\rangle_{\psi(0)}} \tag{10}$$

For the JC-LQB, inserting $H_{\text{L}}$ into Eq. (10) yields its QSL. By inserting Eq. (1) into Eq. (10), we can calculate the QSL of the O-NQB. Thus, computing the variances for both cases gives

$$\begin{gathered}\langle\delta H_{\text{L}}\rangle_{\psi(0)}^2 = ng^2 \\ \langle\delta H\rangle_{\psi(0)}^2 = n!\left[\prod_{j=1}^{k} f(j)\right]^2 g^2\end{gathered} \tag{11}$$

Setting $n = k$ and equating the two expressions in Eq. (11), we derive the unbiased form of $f(n)$

$$f(n)=\begin{cases}1 & ,if\ n=1\\ \dfrac{1}{\sqrt{n-1}} & ,if\ n\geq 2\end{cases} \tag{12}$$

The corresponding QSL time is

$$\tau_{\mathrm{QSL}}=\frac{\pi}{2g\sqrt{n}} \tag{13}$$

We first consider the resonant case ($\Delta=0$). Substituting Eq. (12) and Eq. (6) into Eq. (7) and Eq. (8), we obtain the charging performance metrics for both models

(i) JC-LQB case ($\mathrm{n}=\mathrm{k}=1$)

$$\begin{aligned}\Delta E^{\mathrm{L}}(t)&=\omega\sin^2(gt)\\ \langle P^{\mathrm{L}}(t)\rangle&=\frac{\omega\sin^2(gt)}{t}\end{aligned} \tag{14}$$

The charging time (as defined earlier) is $t_f^{\mathrm{L}}=\frac{\pi}{2g}$.

(ii) O-NQB case ($n=k\geq 2$)

$$\begin{aligned}\Delta E(t)&=n\omega\sin^2\left(g\sqrt{n}t\right)\\ \langle P(t)\rangle&=\frac{n\omega\sin^2\left(g\sqrt{n}t\right)}{t}\end{aligned} \tag{15}$$

Here, the charging time is $t_f=\frac{\pi}{2g\sqrt{n}}\equiv\tau_{\mathrm{QSL}}$, indicating that the nonlinear interaction allows the charging time to saturate the QSL.

The average charging powers over the full process for both cases are

$$\begin{aligned}\langle P^{\mathrm{L}}\left(t_f^{\mathrm{L}}\right)\rangle&=\frac{2g\omega}{\pi}\\ \langle P\left(t_f\right)\rangle&=\frac{2g\omega}{\pi}n^{1.5}\propto n^{1.5}\end{aligned} \tag{16}$$

From Eq. (16), it can be seen that the average charging power of the JC-LQB offers no quantum advantage; however, the average charging power of the O-NQB scales superlinearly with $n$. Thus, our proposed O-NQB exhibits quantum advantage [19]. Accordingly, with the unbiased form of $f(n)$, we have upheld the fairness of this comparison, which further confirms that the introduction of nonlinear interactions effectively enhances the charging performance of QB.

In summary, the O-NQB constructed in this work and the model in Ref.27 can both saturate the QSL in charging time, both possess genuine quantum advantage, and both fall into the research scope of nonlinear quantum batteries. Specifically, the model in Ref.27 takes two bosonic harmonic oscillators as the charger and the battery respectively, and realizes nonlinear coupling through anharmonic interactions containing high-order operators. By contrast, the model in this work adopts a two-level atom as the battery and an optical cavity as the charger, and achieves nonlinear coupling between the two via the multiphoton absorption process, which falls into the research direction of nonlinear light-matter interaction. There are essential differences between the two models in terms of their core architecture and the underlying physical mechanism of their quantum advantage.

**4. Physical mechanism of the quantum advantage**

In this section, we briefly discuss the physical mechanism underlying the quantum advantage of the proposed O-NQB. As illustrated in Fig. 2, which presents the schematic diagram of energy exchange in the nonlinear QB, our work focuses on the energy exchange mode between the charger and the QB. We introduce high-order nonlinear interactions into the energy transfer process, which enables the simultaneous absorption of multiple identical photons from the charger by a single battery unit. This realizes the superlinear transfer of energy, and further endows the nonlinear QB with a distinct quantum advantage originating from the collective multiphoton absorption effect.

In contrast, it is worth noting that for other representative models with verified quantum advantage [14,31,49,50], their physical mechanism is mainly based on the collective linear coupling between a single charger and multiple battery units, with the quantum advantage derived from the global charging operation therein. Meanwhile, Gyhm et al. [49] have proven that the global collective charging operation is a necessary but not sufficient condition for achieving quantum advantage. They clarified that for a physically self-consistent QB system, the charging power can only achieve a quadratic enhancement at most relative to the classical linear scaling, thus setting the theoretical upper bound of the charging power scaling in this field. The proposal of our O-NQB model provides a new research pathway for enhancing the charging performance of quantum batteries, and also establishes a research paradigm for the theoretical investigation of nonlinear interactions between chargers and quantum batteries.

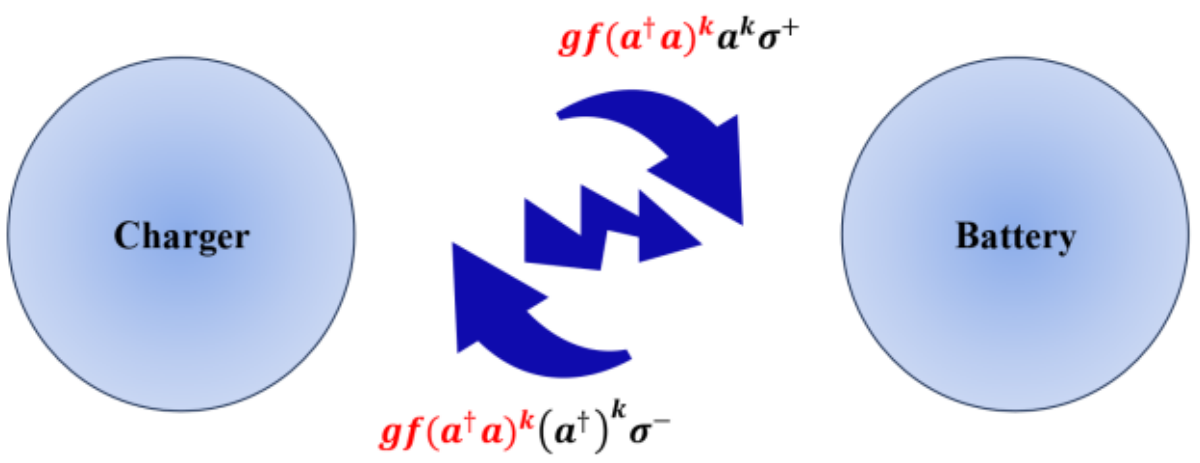

FIG. 2. Schematic diagram of the energy exchange between the charger and the quantum battery, where $g$ is the coupling constant, $f(a^\dagger a)^k$ is the nonlinear function, $a^\dagger(a)$ is the creation (annihilation) operator of the optical field in the charger, and $\sigma^+(\sigma^-)$ is the raising (lowering) operator of the quantum battery.

Furthermore, by analyzing the evolution law of the system's quantum resources throughout the entire charging process, we further clarify that the quantum advantage realized by our model is a genuine quantum advantage. In this work, we adopt the Relative Entropy of Coherence $C_{\text{rel}}(\rho)$ to quantify the dynamic evolution of quantum resources (quantum coherence) in the model. This metric is a widely recognized, standard and robust measure in the field of quantum information, which satisfies all the axiomatic conditions for a coherence measure and can accurately distinguish between quantum processes and classical processes. Its specific definition is given by

$$C_{\text{rel}}(\rho) = S(\rho_\Delta) - S(\rho) \tag{17}$$

where $S(\rho) = -\text{Tr}(\rho \ln \rho)$ is the von Neumann entropy corresponding to the density matrix of the model $\rho = |\psi_I(t)\rangle\langle\psi_I(t)|$, and $\rho_\Delta = \text{diag}(\rho_{00}, \rho_{11})$ denotes the diagonal matrix of the density matrix $\rho$ in the computational basis.

With the photon number fixed, we investigate the time evolution of the stored energy $\Delta E$ of the battery and the relative entropy of coherence $C_{\text{rel}}$ of the system under different detuning conditions, with the results presented in Fig. 3.

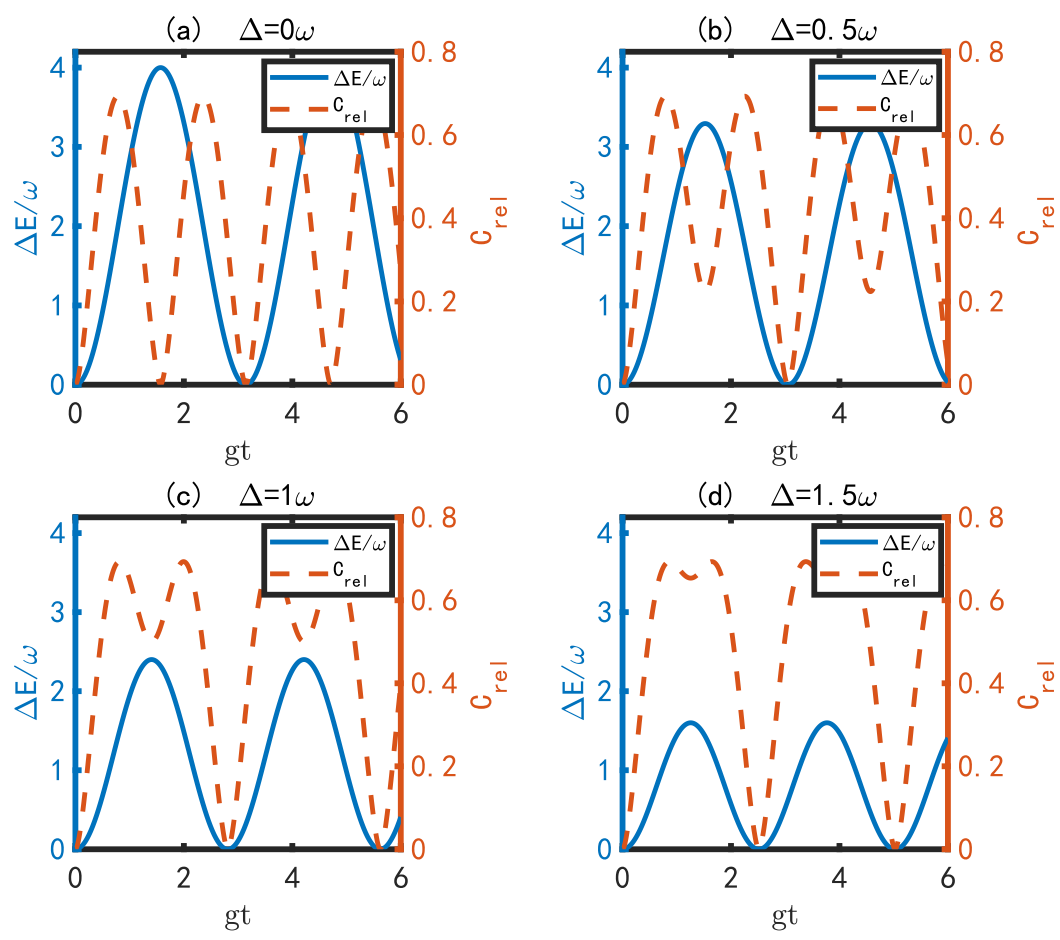

FIG. 3 Time evolution of the stored energy $\Delta E$ and the relative entropy of coherence $C_{\text{rel}}$ for the photon number $n=4$. (a) $\Delta=0\omega$, (b) $\Delta=0.5\omega$, (c) $\Delta=1\omega$, (d) $\Delta=1.5\omega$. The calculation parameter is set as $g=0.5\omega$.

The evolution law of the physical process can be clearly observed from Fig. 3. In the initial stage of charging, the coupling between the battery and the charger is activated, and the coherence of the system increases continuously, corresponding to the continuous accumulation of available quantum resources in the system. The coherence begins to decrease after reaching its maximum value $C_{\text{rel}}^{\max}$ for the first time, and in this stage, the quantum resources of the system are gradually converted into the energy stored in the QB. When the stored energy $\Delta E$ reaches its first peak, the coherence $C_{\text{rel}}$ stops decreasing synchronously, which corresponds to the termination of the charging process. The residual value of coherence at this point is the untransformed remaining quantum resources $C_{\text{rel}}^{\text{rem}}$ in the model. Further analysis reveals that $\Delta E$ and $C_{\text{rel}}$ show a synchronous upward trend in the initial charging stage, whose physical origin is as follows: the initial growth of the stored energy mainly comes from the contribution of incoherent work, while in the subsequent stage where the coherence decreases but the stored energy continues to rise, the energy growth is entirely derived from the conversion contribution of the system's quantum resources. In addition, Fig. 4 shows the effect of detuning on the evolution of $C_{\text{rel}}$ of the system during the charging process for different photon numbers.

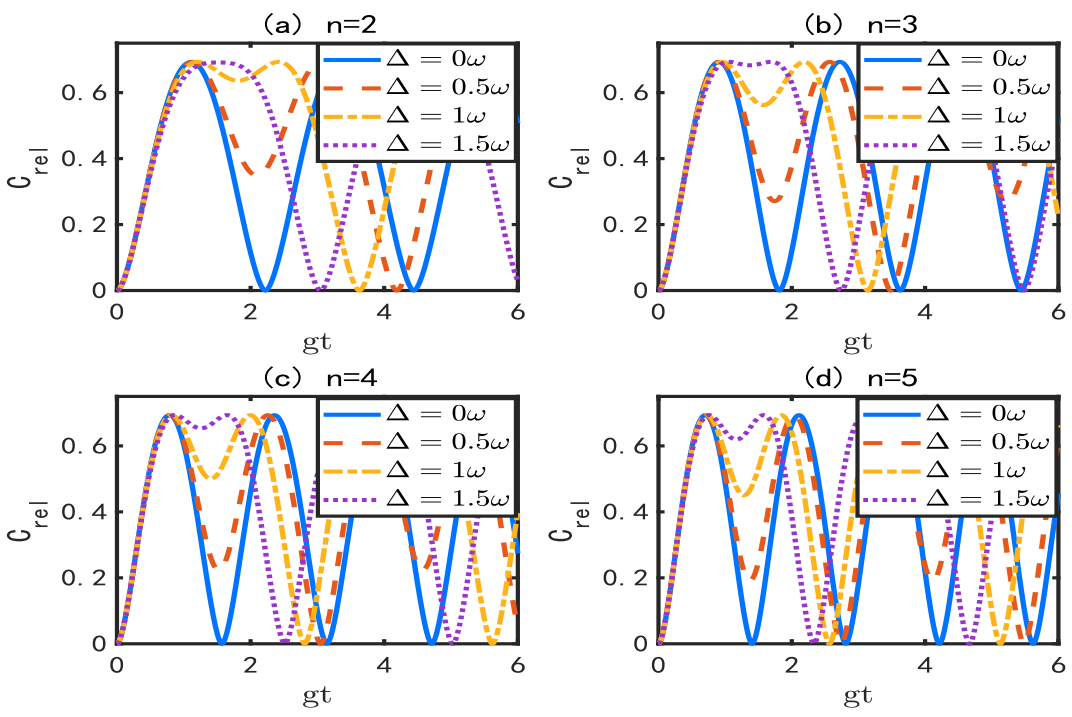

FIG. 4 Evolution of the relative entropy of coherence $C_{\text{rel}}$ under different detuning $\Delta$ during the charging process. (a) $n=2$, (b) $n=3$, (c) $n=4$, (d) $n=5$. The calculation parameter is set as $g=0.5\omega$.

From the results presented in Fig. 3 and Fig. 4, it can be clearly observed that the effective utilization of quantum coherence of the system (defined as $C_{\rm rel}^{\rm max} - C_{\rm rel}^{\rm rem}$, i.e., the difference between the peak value of coherence and the residual coherence at the termination of the charging process) decreases significantly with increasing detuning. From the perspective of quantum resource utilization, this result directly reveals the physical mechanism underlying the detrimental effect of detuning on energy storage enhancement: detuning reduces the conversion efficiency of the system's quantum resources into the energy stored in the battery. Meanwhile, we also find that with the increase of the photon number $n$, the effective utilization of the system's quantum coherence is still significantly enhanced even under large detuning conditions.

In this section, from the perspective of quantum coherence, a core quantum resource, we rigorously demonstrate that the genuine quantum advantage arises from multiphoton absorption. This O-NQB can effectively enhance the performance of quantum batteries.

## 5. Performance of the nonlinear quantum battery

In Section 3 derived the fair form of the nonlinear function, while briefly discussing the stored energy $\Delta E(t)$, charging time $t_f$, and the full-process average charging power $\langle P(t_f)\rangle$ of the O-NQB under this form. Next, we elaborate on these charging performance metrics for the resonant case. As shown in Fig. 5 (a) illustrates the time evolution of stored energy, (b) depicts the relationship between charging time and the coupling constant, (c) presents the correlation between the full-process average charging power and the coupling constant.

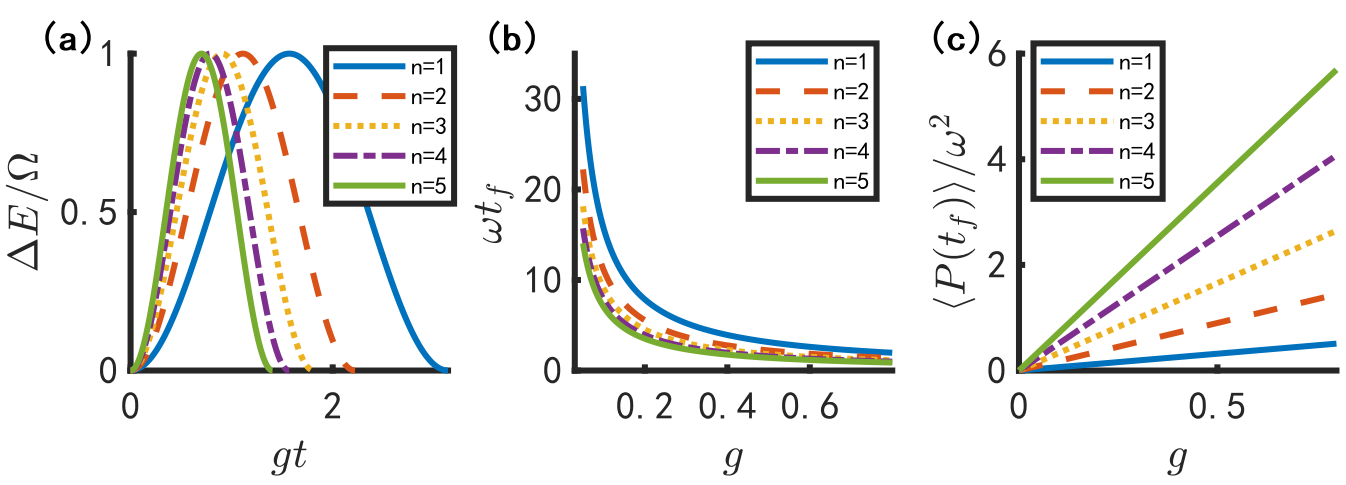

FIG. 5. For the resonant case ($\Delta = 0$), performance of the O-NQB, (a) Time evolution of $\Delta E(t)$, (b) Relationship between $t_f$ and $g$ for different photon numbers, (c) Relationship between $\langle P(t_f)\rangle$ and $g$.

From Fig. 5(a), the time for the stored energy to reach its peak decreases with the increase of the photon number $n$, which means the charging process is accelerated. Additionally, Fig. 5(b) shows that charging speed accelerates with an increase in $g$. In Fig. 5(c), we further observe that the full-process average charging power is positively correlated with both the coupling constant and photon number. Moreover, when $n = 1$ (solid blue line), the curve corresponds to the JC-LQB case; when $n \geq 2$, it corresponds to the O-NQB case. Fig. 5 clearly shows that the charging performance of the nonlinear curves is significantly superior to the blue (linear case) curve, demonstrating that our O-NQB model indeed improves charging performance and exhibits quantum advantage.

Notably, Ref.27 only investigated energy storage and charging power, focusing on the effects of Fock states, coherent states, and squeezed vacuum states on charging performance, and proved that the Fock state is the optimal initial state. In contrast, in this section we will perform a comprehensive full-dimensional characterization of the charging performance of the O-NQB: beyond energy storage and charging power, we will further examine charging precision, extractable work (ergotropy [3]), as well as other critical charging performance metrics.

For a quantum battery to supply energy stably in practical applications, high charging precision (i.e., small quantum energy fluctuations during charging) is essential. We define the quantum energy fluctuation $\Sigma^2$ as

$$\Sigma^2 = \mathrm{Tr}[\rho_{\mathrm{B}}(t)H_{\mathrm{B}}^2] - (\mathrm{Tr}[\rho_{\mathrm{B}}(t)H_{\mathrm{B}}])^2 \tag{18}$$

Substituting Eq. (6) and Eq. (12) into Eq. (18) yields the form of $\Sigma^2$ for the O-NQB case

$$\Sigma^2 = \Omega^2|c_1(t)|^2|c_2(t)|^2 \tag{19}$$

This result indicates that the battery's energy fluctuation depends on the excited-state probability $|c_2(t)|^2$, ground-state probability $|c_1(t)|^2$, and the energy level spacing $\Omega$ of the two-level system. The fluctuation is 0 when $|c_1(t)|^2 = 0$ or 1, and maximized when $|c_1(t)|^2 = 0.5$. For $\Delta = 0$ and at the end of charging $t = t_f$, $\Sigma^2(t_f) = 0$. Thus, our nonlinear quantum battery achieves high charging precision at the termination of charging.

Additionally, the extractable energy of the O-NQB (termed ergotropy [3]) is another critical performance metric. Next, the ergotropy is introduced in detail. After charging is completed, the density matrix of the O-NQB we obtain is $\rho_{\mathrm{B}}(t)$. It is assumed that the system can be fully controlled: any unitary evolution can be generated by applying an appropriate control field to the system; after the transformation is completed, the control field is turned off. Under the action of the optimal unitary transformation $\widehat{\mathrm{U}}$, the state transformation takes the form $\rho_{\mathrm{B}} \to \sigma = \widehat{\mathrm{U}}\rho_{\mathrm{B}}\widehat{\mathrm{U}}^\dagger$. We refer to $\sigma$ as the passive state [51] of the O-NQB, and no energy can be extracted from the passive state through any unitary transformation. Therefore, the ergotropy can be expressed as

$$\mathcal{E}(\rho_{\mathrm{B}}) = \max_U W(\rho_{\mathrm{B}}, U) = \mathrm{Tr}(\rho_{\mathrm{B}}H_{\mathrm{B}}) - \mathrm{Tr}\left(\widehat{U}\rho_{\mathrm{B}}\widehat{U}^\dagger H_{\mathrm{B}}\right) = \sum_k \varepsilon_k(\rho_{kk} - r_k) \tag{20}$$

where, after sorting the labels of the eigenstates of $H_{\mathrm{B}}$ and $\rho_{\mathrm{B}}$, we have $H_{\mathrm{B}} = \sum_{k=1}^{d} \varepsilon_k |\varepsilon_k\rangle\langle\varepsilon_k|$ with $\varepsilon_k \le \varepsilon_{k+1}$, and $\rho_{\mathrm{B}} = \sum_{k=1}^{d} r_k |r_k\rangle\langle r_k|$, with $r_k \ge r_{k+1}$. Meanwhile, $\sigma = \sum_{k=1}^{d} r_k\, |\varepsilon_k\rangle\langle\varepsilon_k|$, and $\rho_{kk}$ is the population of $\rho_{\mathrm{B}}$ in the *k*-th energy eigenstate, which can be expressed as $\rho_{kk} = \sum_{k'} r_{k'} |\langle r_{k'}|\varepsilon_k\rangle|^2$.

Substituting Eq. (6) and Eq. (12) into Eq. (20), the ergotropy of this O-NQB takes the form

$$\mathcal{E}(\rho_{\mathrm{B}}) = \Omega(2|c_2(t)|^2 - 1)\Theta\left(|c_2(t)|^2 - \frac{1}{2}\right) \tag{21}$$

where, $\Theta(x - x_0)$ is the Heaviside function, which satisfies $\Theta(x - x_0) = 0$ for $x < x_0$, $\Theta(x - x_0) = \frac{1}{2}$ for $x = x_0$, $\Theta(x - x_0) = 1$ for $x > x_0$. For the resonant case ($\Delta = 0$), Fig. 6(a) shows the time evolution of $\mathcal{E}(\rho_{\mathrm{B}})$, where the extractable work for $n = 1$ (solid blue line, JC-LQB) is lower than that of the nonlinear battery. This confirms that the O-NQB enhances performance, with this advantage stemming from multi-photon absorption effects.

Finally, we discuss the battery's efficiency, defined as the ratio of extractable work to stored energy. The efficiency

$$\eta = \frac{\mathcal{E}(\rho_{\mathrm{B}})}{\Delta E(t)} \tag{22}$$

Incorporate Eq. (15) and Eq. (21) into Eq. (22), the efficiency becomes

$$\eta = \begin{cases} 0 & , |c_2(t)|^2 \le \dfrac{1}{2} \\ 2 - \dfrac{1}{|c_2(t)|^2} & , |c_2(t)|^2 > \dfrac{1}{2} \end{cases} \tag{23}$$

For $\Delta = 0$, Fig. 6(b) illustrates the time evolution of $\eta$: the efficiency reaches 100% in all cases (a consequence of the model being a closed system). Additionally, the time to reach peak efficiency shortens as $n$ increases, indirectly confirming that nonlinear interactions accelerate both charging and work extraction.

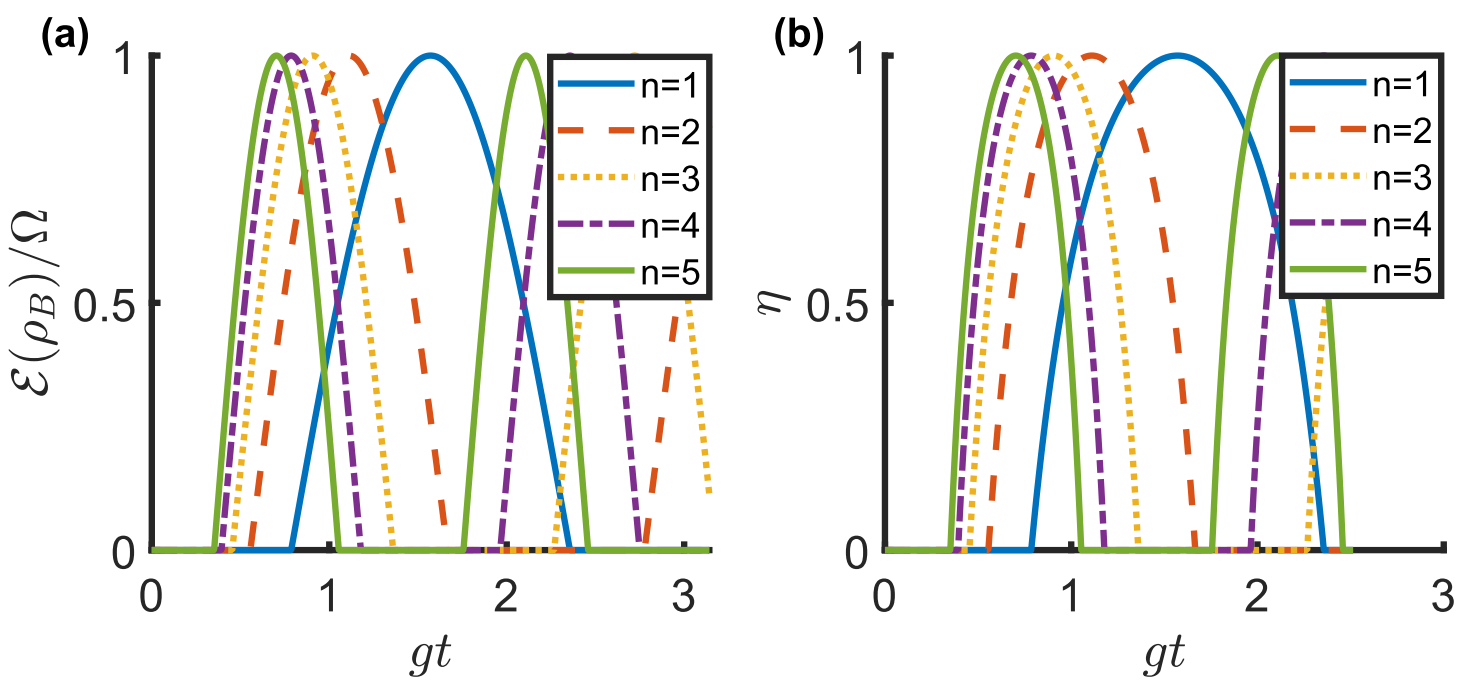

FIG. 6. For the resonant case $\Delta = 0$, performance of the O-NQB for different photon numbers, (a) Time evolution of $\mathcal{E}(\rho_{\mathrm{B}})$, (b) Time evolution of $\eta$.

All prior discussions focused on the resonant case. We now investigate the effect of detuning (moving from $\Delta = 0$ to $\omega$) on the O-NQB's performance (using the solution method detailed earlier). First, we examine detuning's impact on maximum stored energy, it is straightforward to see $\Delta E(t) = \Omega|c_2(t)|^2$, so the charging time is $t_f = \frac{\pi}{R(\Delta)}$, and $\Delta E(t_f)$ is the maximum stored energy. Fig. 7 details how detuning $\Delta$ affects $\Delta E(t_f)$, showing that with detuning present, maximum stored energy increases with $n$, but larger detuning is detrimental to energy storage.

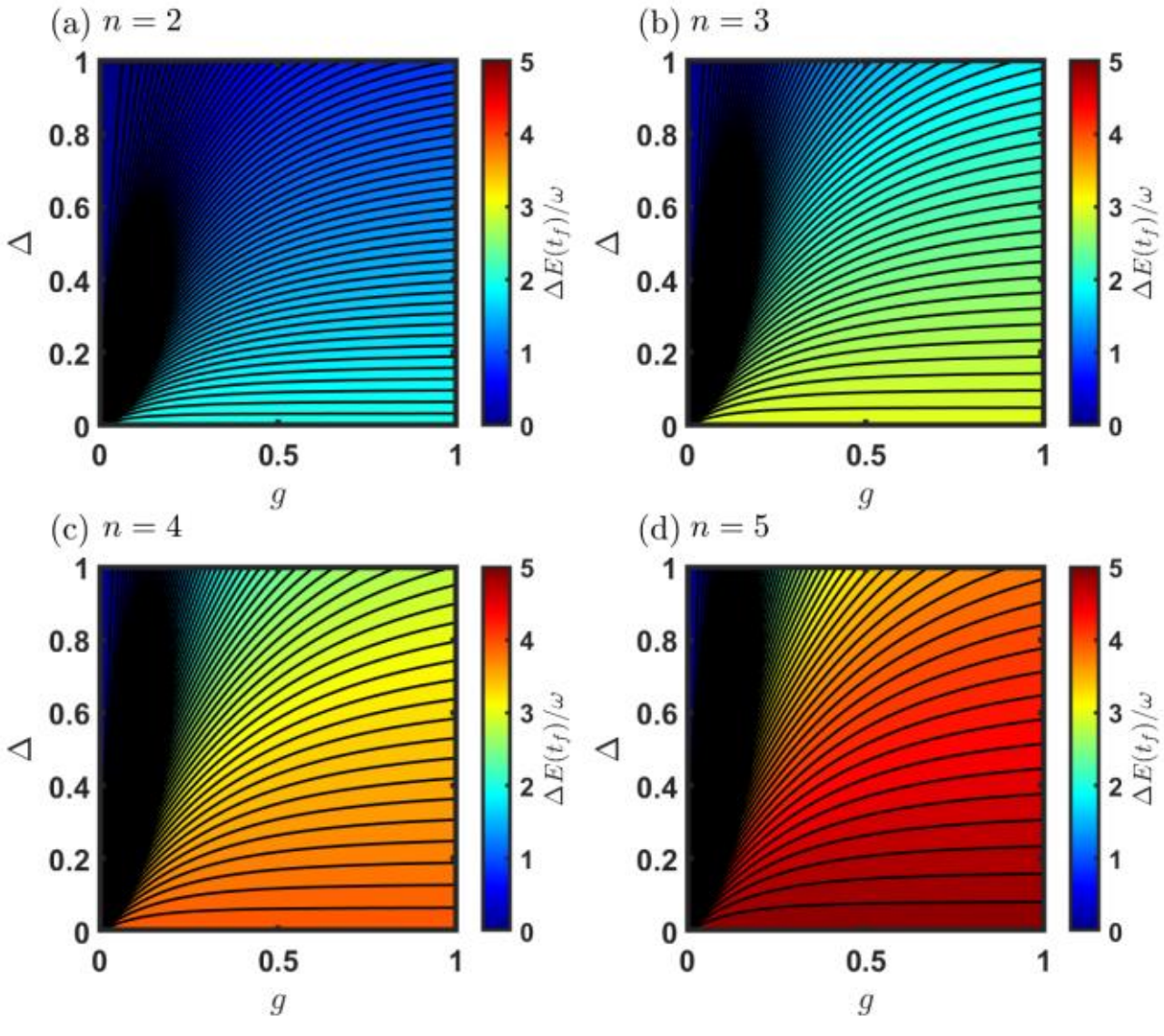

FIG. 7. Effect of detuning $\Delta$ and coupling constant $g$ on the maximum stored energy $\Delta E(t_f)$ of the O-NQB for different photon numbers (a) $n = 2$, (b) $n = 3$, (c) $n = 4$, (d) $n = 5$, $(\Delta = 0 \rightarrow \omega)$.

In Fig. 8 plots $\Delta$ effect on $t_f$, revealing that as $g$ increases, $\Delta$ impact on $t_f$ weakens. In the weak coupling regime, increasing $\Delta$ shortens $t_f$ (strong acceleration), this effect is less pronounced in the strong coupling regime. Additionally, charging accelerates significantly as $g$ increases.

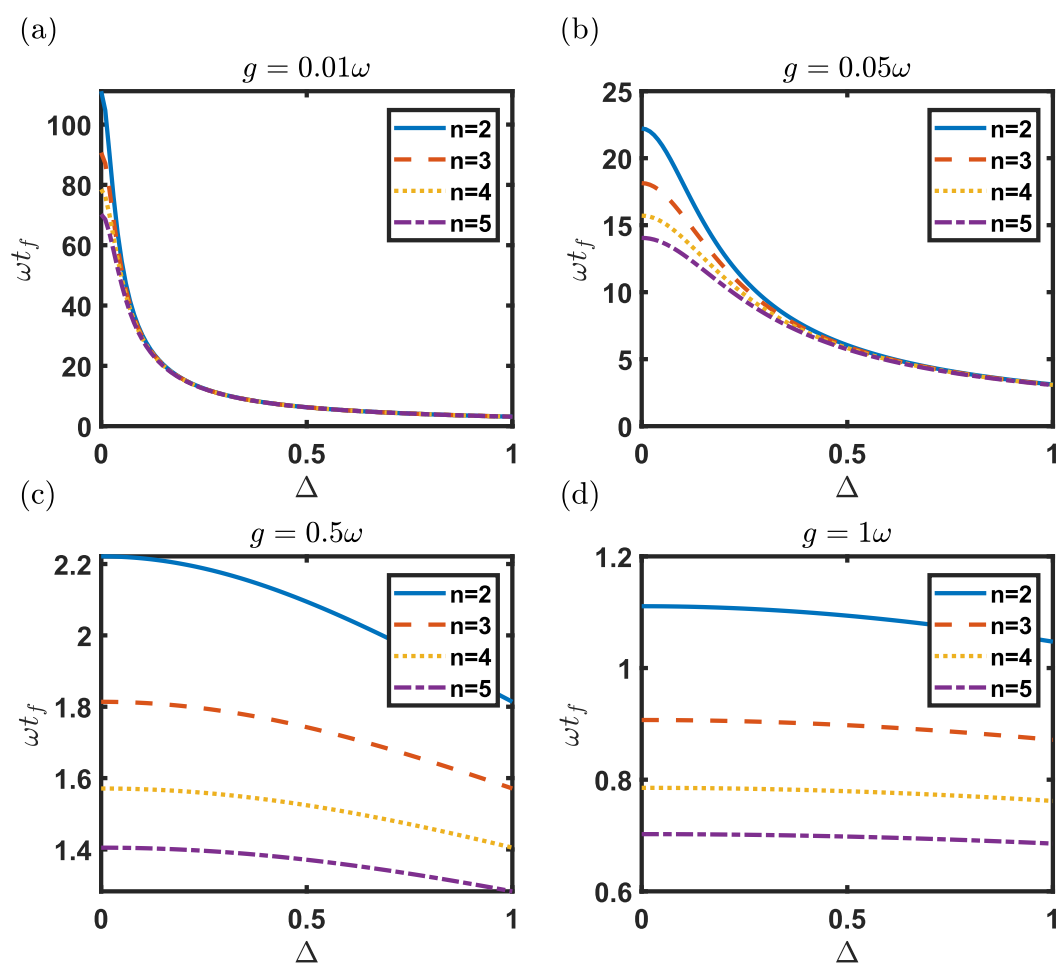

FIG. 8. Effect of detuning $\Delta$ on the charging time $t_f$ for different coupling constants, (a) $g = 0.01\omega$, (b) $g = 0.05\omega$, (c) $g = 0.5\omega$, (d) $g = \omega$, $(\Delta = 0 \to \omega)$.

Fig. 9 explores detuning's effect on $\langle P(t_f)\rangle$ (calculated via Eq. (6) and Eq. (8) as $\langle P(t_f)\rangle = \frac{\Omega R(\Delta)|c_2(t_f)|^2}{\pi}$). Detuning harms the full-process average charging power, but $\langle P(t_f)\rangle$ increases with $g$ and $n$. Calculations confirm $\langle P(t_f)\rangle \sim n^{1.5}$, thus the quantum advantage exists.

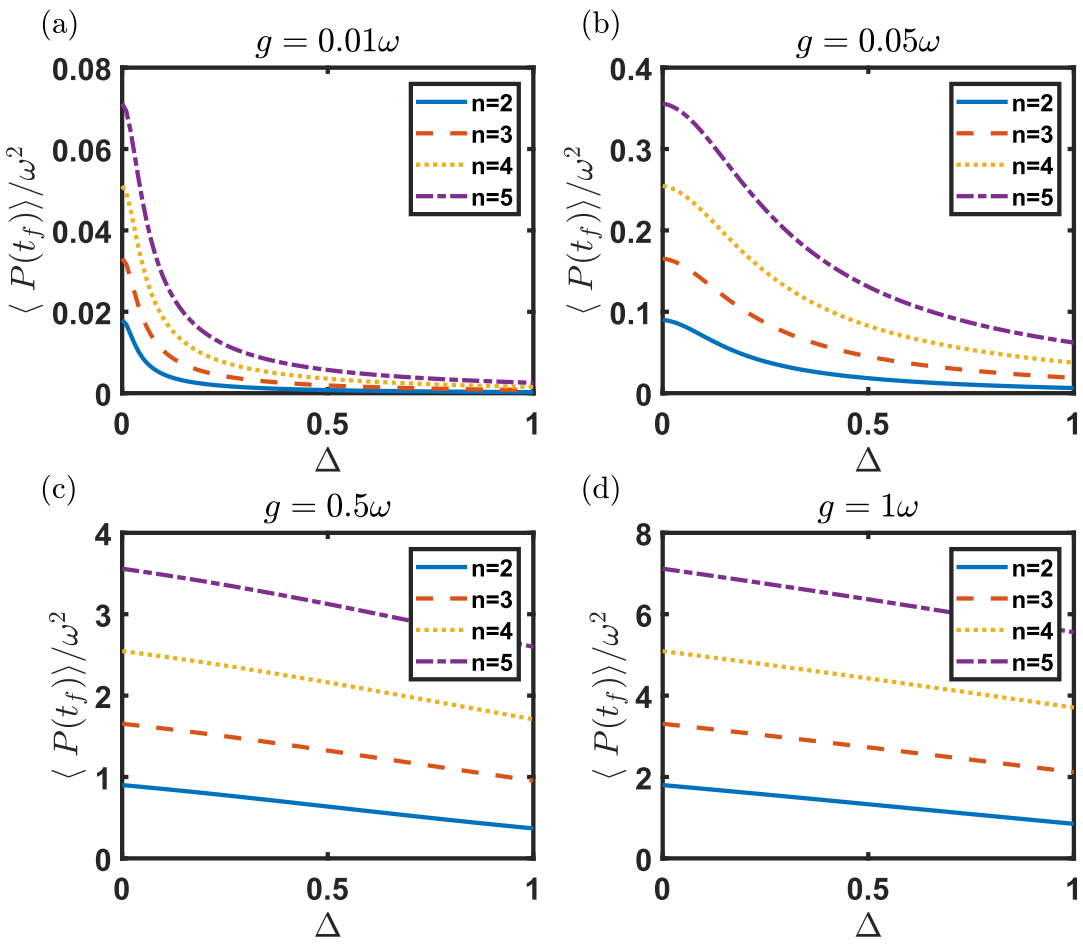

FIG. 9. Effect of detuning $\Delta$ on $\langle P(t_f)\rangle$ for different coupling constants (a) $g = 0.01\omega$, (b) $g = 0.05\omega$, (c) $g = 0.5\omega$, (d) $g = \omega$, $(\Delta = 0 \to \omega)$.

By combining Eq. (6) and Eq. (19), we examine detuning's effect on charging precision at the end of charging $\Sigma^2(t_f)$. From Fig. 10 with detuning, stable charging precision is hard to maintain. In the strong coupling regime, $\Sigma^2(t_f)$ increases with $\Delta$ and $n$. In the weak coupling regime, precision first rises then falls. Charging precision also improves with $g$.

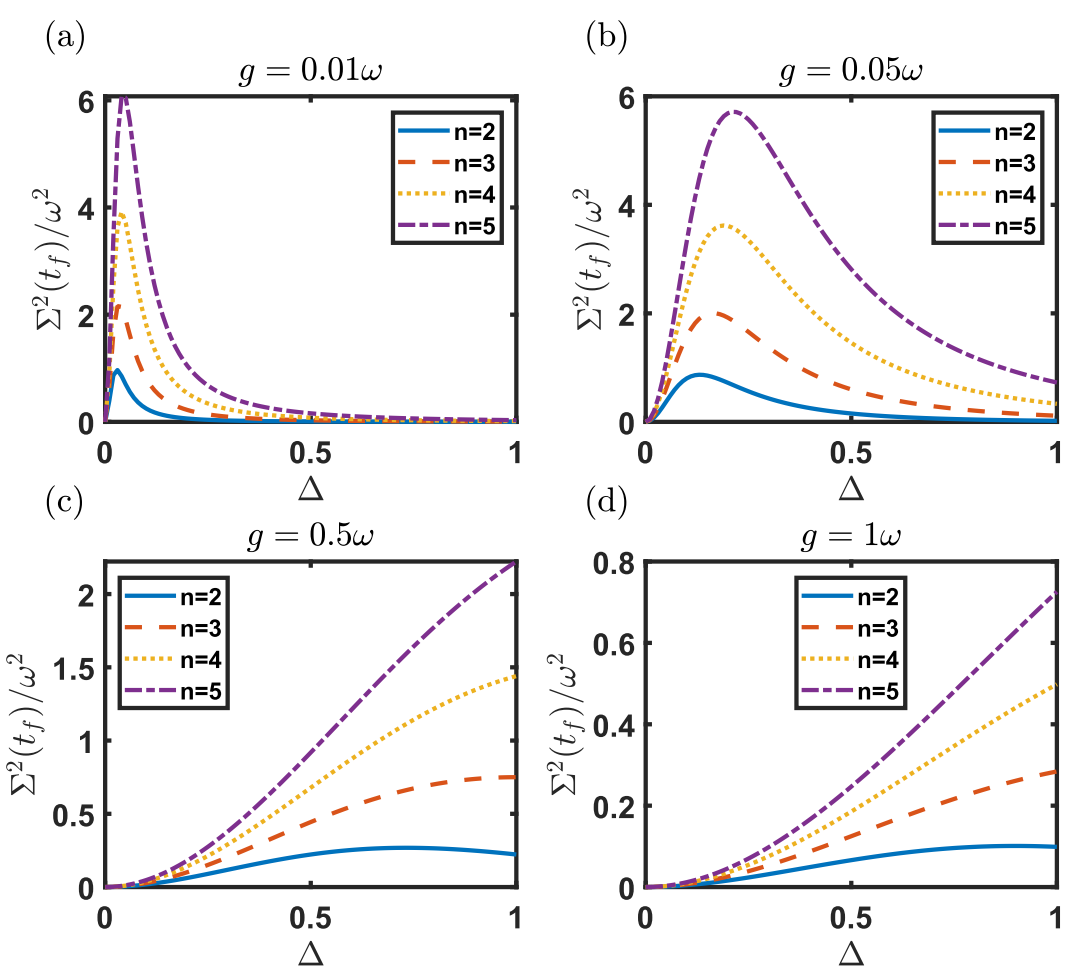

FIG. 10. Effect of detuning $\Delta$ on the quantum energy fluctuation $\Sigma^2(t_f)$ at the end of charging for different coupling constants (a) $g = 0.01\omega$, (b) $g = 0.05\omega$, (c) $g = 0.5\omega$, (d) $g = \omega$, $(\Delta = 0 \rightarrow \omega)$.

Finally, we use Eq. (21) and Eq. (23) to investigate detuning's effect on $\mathcal{E}_{\max}$ (maximum extractable work) and $\eta_{\max}$ (maximum efficiency)—key efficiency metrics. $\mathcal{E}_{\max}$ is the extractable energy of $\rho_{\mathrm{B}}(t_f)$ (the battery's density matrix at charging termination)

$$\mathcal{E}_{\max} = \Omega\left(2\left|c_2(t_f)\right|^2 - 1\right)\Theta\left(\left|c_2(t_f)\right|^2 - \frac{1}{2}\right) \tag{24}$$

And, $\eta_{\max}$ is the ratio of $\mathcal{E}_{\max}$ to the maximum stored energy $E_{\max} = \Delta E(t_f)$, with the form

$$\eta_{\max} = \begin{cases} 0 & ,\left|c_2(t_f)\right|^2 \leq \frac{1}{2} \\ 2 - \frac{1}{\left|c_2(t_f)\right|^2} & ,\left|c_2(t_f)\right|^2 > \frac{1}{2} \end{cases} \tag{25}$$

Fig. 11 shows detuning's effect on $\mathcal{E}_{\max}$, detuning harms maximum ergotropy, and in the weak coupling regime, $\mathcal{E}_{\max}$ drops rapidly to 0 for small $\Delta$.

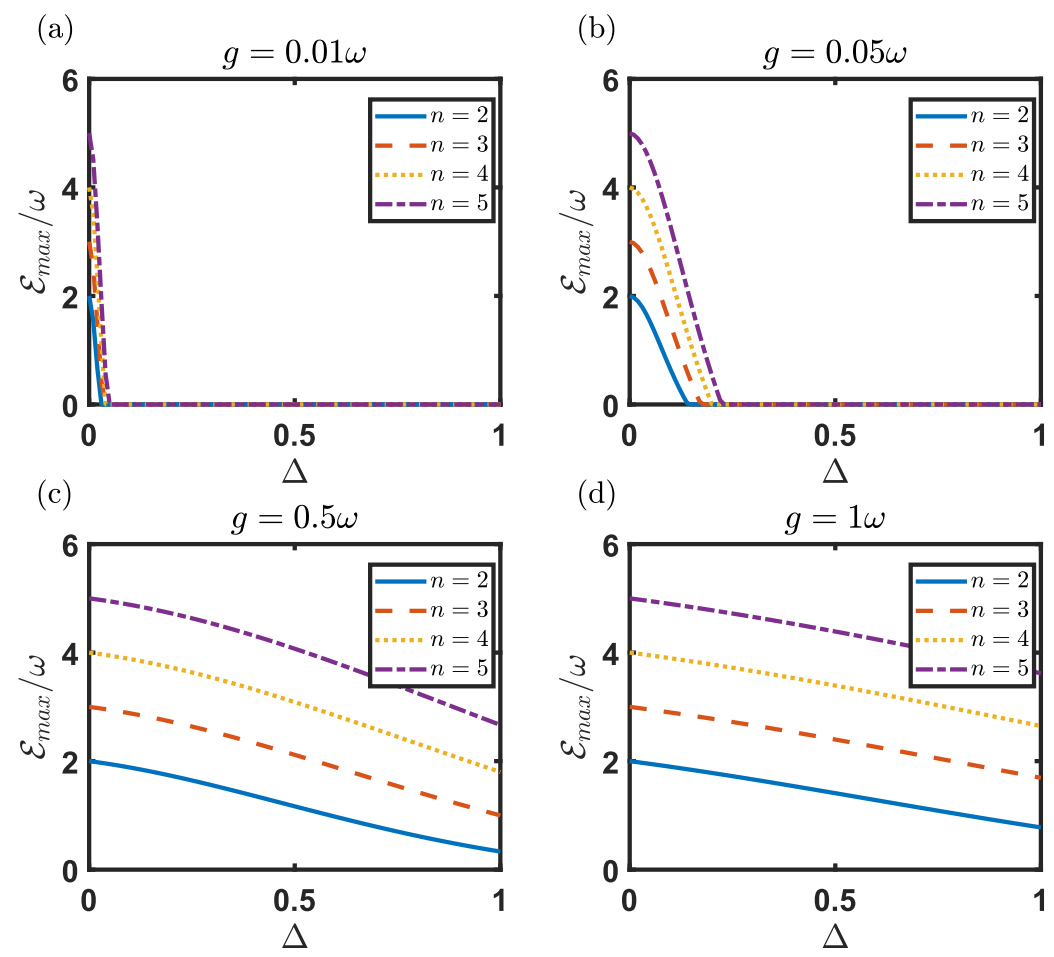

FIG. 11. Effect of detuning $\Delta$ on the maximum ergotropy $\mathcal{E}_{\max}$ for different coupling constants (a) $g = 0.01\omega$, (b) $g = 0.05\omega$, (c) $g = 0.5\omega$, (d) $g = \omega$, $(\Delta = 0 \rightarrow \omega)$.

In Fig. 12 shows detuning's effect on $\eta_{\max}$, $\eta_{\max}$ decreases with $\Delta$, and in the weak coupling regime, it drops rapidly to 0 for small $\Delta$. However, in the strong coupling regime (as $g$ increases from $0.5\omega$ to $\omega$), $\eta_{\max}$ improves.

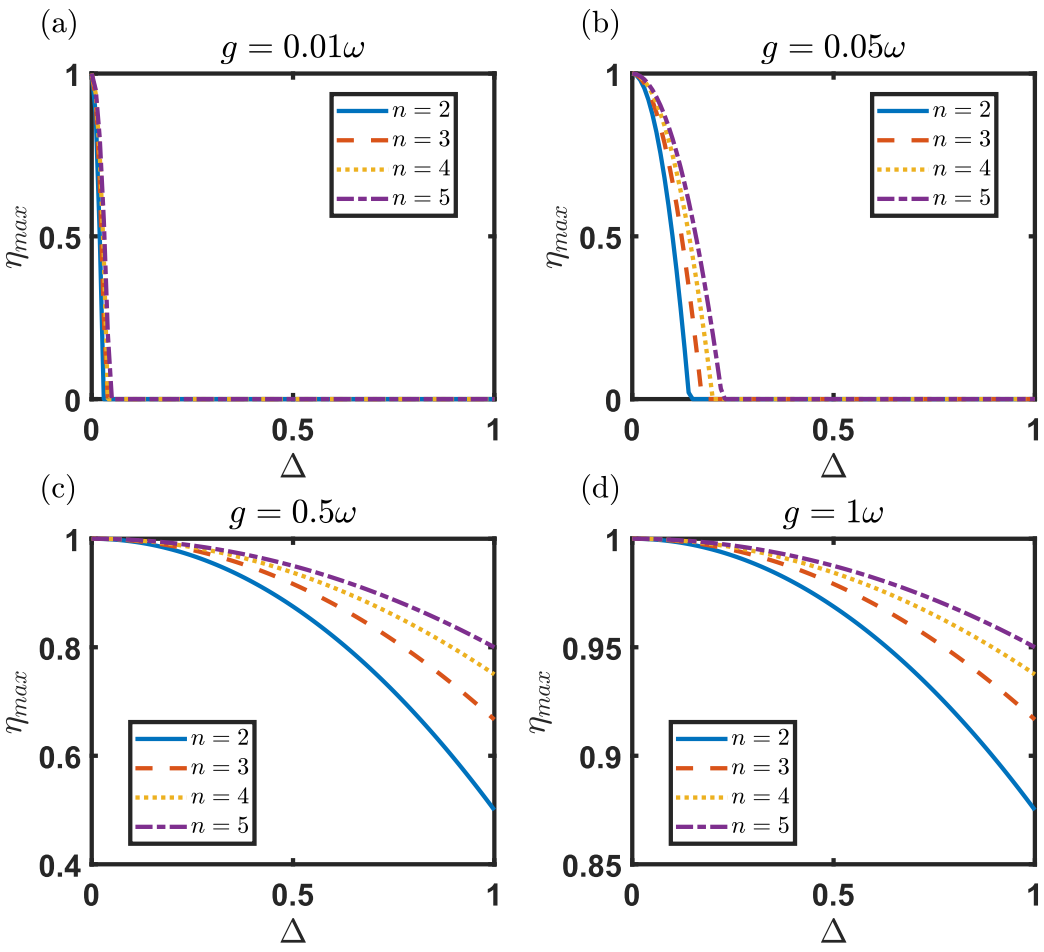

FIG. 12. Effect of detuning $\Delta$ on the maximum efficiency $\eta_{\max}$ for different coupling constants (a) $g = 0.01\omega$, (b) $g = 0.05\omega$, (c) $g = 0.5\omega$, (d) $g = \omega$, ($\Delta = 0 \rightarrow \omega$).

In summary, this section has investigated the O-NQB's performance under the derived unbiased optical-field-dependent nonlinear interaction function. For the resonant case, nonlinear interactions enhance all charging metrics relative to the linear battery, with the full-process average charging power exhibiting a distinct superlinear quantum advantage. And we also found that detuning can accelerate charging but has adverse effects on other performance metrics.

Finally, the above discussions on the influence of detuning on the charging performance of the O-NQB have been limited to the presentation of numerical simulations, and it is necessary to provide a physical interpretation of the underlying mechanism for the effect of detuning on the charging performance of this O-NQB. The energy level formula can be readily obtained as

$$E_n = \left(n\omega + \frac{\Omega}{2}\right) \pm R_n(\Delta) = \frac{3n}{2}\omega \pm \frac{1}{2}\sqrt{4g^2 n + \Delta^2} \tag{26}$$

Based on the above equation, the influence of detuning on the energy levels is obtained via numerical simulation, which is presented in Fig. 13 (a). It can be seen from the figure that the energy level splitting becomes increasingly pronounced as the detuning increases.

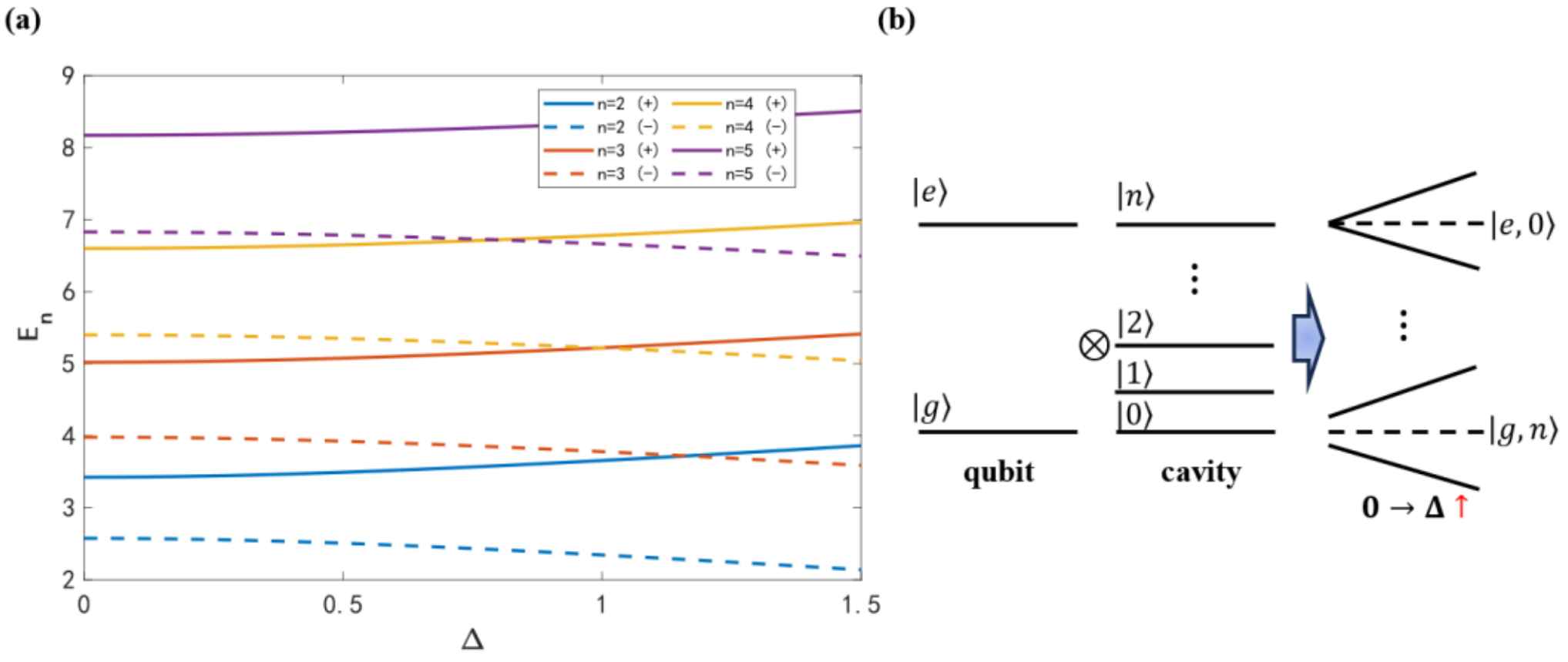

FIG. 13 (a) Variation of the energy spectrum with the detuning $\Delta$, where $g = 0.3\omega$. (b) For the energy level diagram of the O-NQB model, the energy level spacing widens as the $\Delta$ increases from left to right.

When the detuning is much larger than the coupling strength, $E_n^{\pm} \approx \frac{3n}{2}\omega \pm \frac{1}{2}\Delta \pm \frac{g^2 n}{\Delta}$, where the energy level splitting is dominated by the detuning, the schematic diagram illustrating the effect of detuning on the energy levels is presented in Fig. 13 (b). In addition, the probability of the excited state can be calculated as $|c_2(t)|^2 = \frac{4g^2\xi^2}{4g^2\xi^2+\Delta^2}\sin^2\left(\frac{\sqrt{4g^2\xi^2+\Delta^2}}{2}t\right)$. It can be found that its amplitude decreases with the increase of detuning, and the effective transition strength is significantly suppressed. Furthermore, as shown in Fig. 3 and Fig. 4, the effective utilization of quantum coherence also decreases as the detuning increases. We provide a physical interpretation of this underlying mechanism from two perspectives (the energy level variation of the system and the evolution of quantum resources during the charging process).

**6. Design of the physical implementation scheme**

In theory, any physical platform capable of realizing quantum computers can be used to implement QBs. Moreover, exploring applications of QB is critical. For instance, using them to power quantum computers [19], though this remains a conceptual idea. Building on this vision, this paper employs superconducting quantum circuits to construct the proposed O-NQB, further promoting the realization of QB that power superconducting quantum computers.

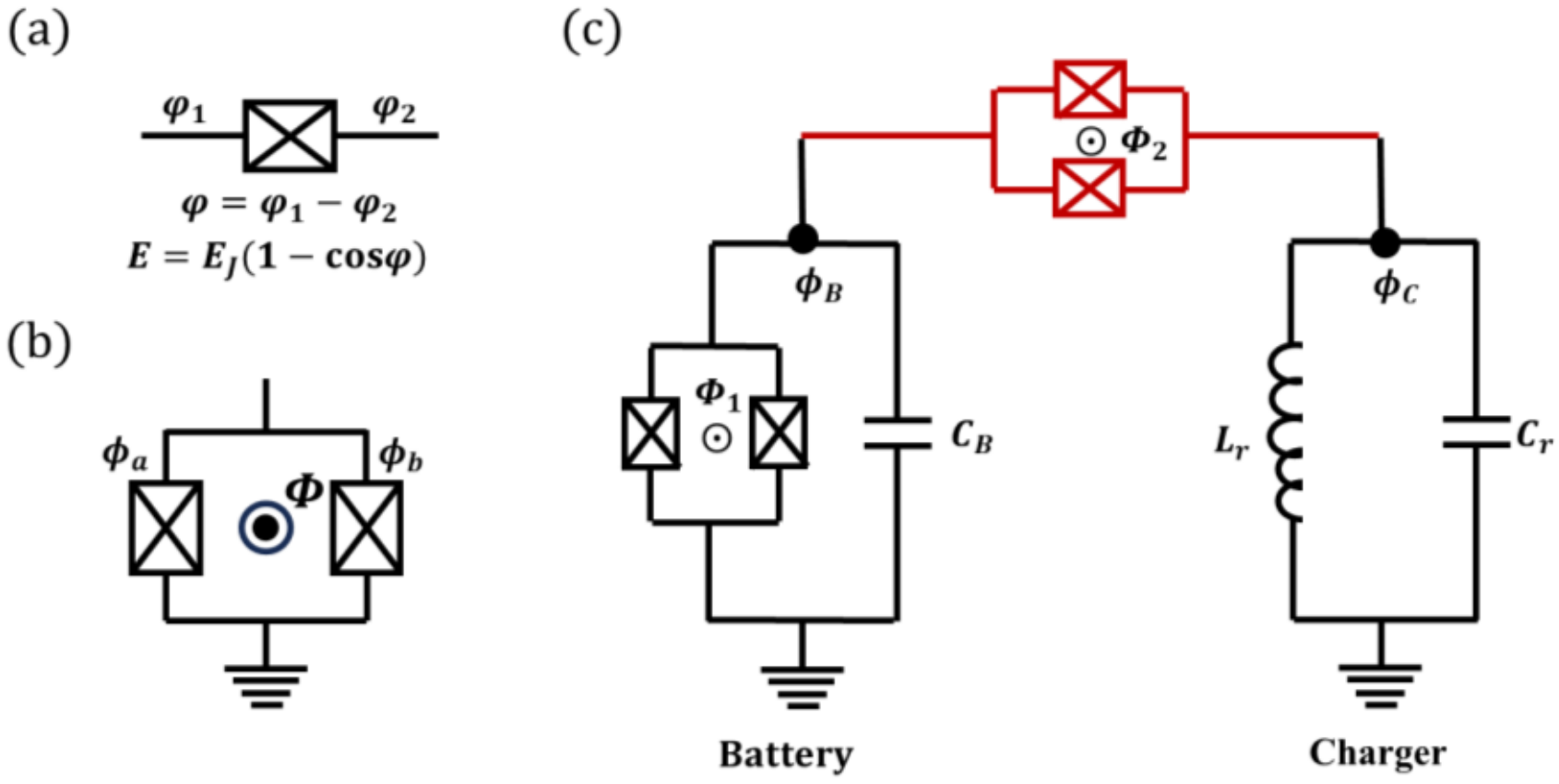

FIG. 14. (a) Equivalent circuit structure of a Josephson junction; $\varphi$ is the superconducting phase difference across the Josephson junction, $E_J$ is the critical energy parameter of the Josephson junction (an intrinsic property of the junction), and $E$ is the energy of the Josephson junction. (b) Equivalent circuit structure of a superconducting quantum interference device (SQUID); $\phi_a$ and $\phi_b$ denote the respective phase differences of the two parallel Josephson junctions in the SQUID, and $\Phi$ represents the external magnetic flux through the SQUID. (c) Superconducting quantum circuit implementation scheme for the O-NQB model; here, the charge qubit serves as the battery, the inductor-capacitor resonator (LC resonator) acts as the charger, and they are coupled via a SQUID. Specifically, $\Phi_1$, $\phi_B$, and $C_B$ denote the external magnetic flux through the SQUID in the battery unit, the superconducting quantum phase at the circuit node, and the superconducting capacitor in the circuit, respectively; $\Phi_2$ is the external magnetic flux of the SQUID in the coupling unit; $L_r$, $\phi_C$, and $C_r$ are the inductor, the phase at the circuit node, and the capacitor in the charger unit, respectively.

Superconducting quantum circuit units typically use two charge qubits coupled via capacitors,

but this coupling is linear. To introduce nonlinear interactions, we can replace the capacitor with Josephson junction coupling. In Fig. 14(a), this is because Josephson junctions exhibit nonlinearity, with their Hamiltonian given by $\mathcal{H} = E_J(1-\cos\varphi)$. Thus, the design scheme in this paper (shown in Fig. 14(c)) uses a single charge qubit coupled to an inductor-capacitor resonator (LC resonator) via a superconducting quantum interference device (SQUID).

Fig. 14(b) shows the structure of the SQUID, which consists of two parallel Josephson junctions. SQUID can be controlled via external magnetic fields and can be treated as an equivalent single Josephson junction, as briefly introduced below

$$I_J = I_a + I_b = I_0(\sin\phi_a + \sin\phi_b) = 2I_0\cos\left(\frac{\phi_a-\phi_b}{2}\right)\sin\left(\frac{\phi_a+\phi_b}{2}\right) \tag{27}$$

Here, $I_J$ denotes the total Josephson current of the SQUID, $I_a$ and $I_b$ represent the individual Josephson currents of the two parallel Josephson junctions composing the SQUID, $I_0$ is the critical current of a single Josephson junction, and $\phi_a$ and $\phi_b$ are the respective superconducting phase differences of the two Josephson junctions. Let, $I_C = 2I_0\cos\left(\frac{\phi_a-\phi_b}{2}\right)$ and $\phi = \frac{\phi_a+\phi_b}{2}$, Eq. (27) can then be rewritten as

$$I_J = I_C\sin\phi \tag{28}$$

The critical energy parameter of the Josephson junction is then

$$E_{JSQ} = \frac{\phi_0 I_0}{\pi}\cos\left(\frac{\phi_a-\phi_b}{2}\right) = E_{J,\max}\left|\cos\left(\frac{\pi\Phi}{\phi_0}\right)\right| \tag{29}$$

where, $\phi_a - \phi_b = \frac{2\pi\Phi}{\phi_0}$, and $\phi_0 = \frac{h}{2e}$ (the flux quantum).

Thus, the energy of this SQUID has a form similar to that of a single Josephson junction

$$E_{SQ} = E_{JSQ}(1-\cos\phi) \tag{30}$$

The Hamiltonian of the scheme in Fig. 14(c) is

$$\begin{gathered}\mathcal{H}_{\text{tot}} = \mathcal{H}_{\text{ba}} + \mathcal{H}_{\text{co}} + \mathcal{H}_{\text{ch}}\\ \mathcal{H}_{\text{ba}} = \frac{Q_B^2}{2C_B} + E_{JSQ1}(1-\cos\phi_B)\\ \mathcal{H}_{\text{co}} = E_{JSQ2}[1-\cos(\phi_C-\phi_B)]\\ \mathcal{H}_{\text{ch}} = \frac{Q_C^2}{2C_r} + \frac{\phi_C^2}{2L_r}\end{gathered} \tag{31}$$

Here, $\mathcal{H}_{\text{tot}}$, $\mathcal{H}_{\text{ba}}$, $\mathcal{H}_{\text{co}}$, and $\mathcal{H}_{\text{ch}}$ denote the total Hamiltonian, the battery unit Hamiltonian, the coupling SQUID Hamiltonian, and the charger Hamiltonian, respectively. Additionally, $E_{JSQ1}$ is the Josephson energy of the SQUID in the battery unit, $E_{JSQ2}$ is the Josephson energy of the coupling-unit SQUID, $Q_B$ is the battery unit charge, $Q_C$ is the capacitive charge of the charger unit, $C_r$ and $C_B$ are the capacitances in their respective circuits, $L_r$ is the inductor in the charger, and $\phi_B$ and $\phi_C$ denote their respective the phase. We perform second quantization on each Hamiltonian in Eq. (31) as follows

(i) Second quantization of the battery unit Hamiltonian

Introduce the Cooper pair number $m = \frac{Q_B}{2e}$, magnetic flux $\Phi_B = \frac{2\pi}{\phi_0}\phi_B$, and $E_C = \frac{e^2}{2C_B}$, which satisfy the commutation relation $[\Phi_B, m] = i$. The Hamiltonian can thus be rearranged as

$$\mathcal{H}_{\text{ba}} = 4E_C m^2 + \frac{1}{2}E_{JSQ1}\left(\frac{\phi_0}{2\pi}\right)^2\Phi_B^2 - \frac{1}{24}E_{JSQ1}\left(\frac{\phi_0}{2\pi}\right)^4\Phi_B^4 \tag{32}$$

Next, we introduce the creation and annihilation operators

$$b = \frac{1}{\sqrt{2K}}(\Phi_B + iKm)$$
$$b^\dagger = \frac{1}{\sqrt{2K}}(\Phi_B - iKm) \quad (33)$$
$$K = \sqrt{\frac{8E_C}{E_{JSQ1}}}$$

By discarding the boson non-conservation term, the quantized Hamiltonian is then

$$H_{\text{ba}} = \omega_q b^\dagger b + \frac{\alpha}{2} b^\dagger b^\dagger bb \quad (34)$$

where, $\omega_q = \sqrt{8E_C E_{JSQ1}} - E_C$, $\alpha = -E_C$. We select the two lowest energy levels as the ground and excited states of the two-level system. Numerical simulations show that the energy levels are stable when $E_J \gg E_C$ [52], so the final Hamiltonian is

$$H_{\text{ba}} = \frac{1}{2}\omega_q \sigma_z \quad (35)$$

where, $\omega_q$ is the two-level transition frequency, tunable via the external magnetic field $\Phi_1$. Additionally, from Eq. (33) $\phi_B = \frac{\phi_0}{2\pi}\Phi_B = \frac{\phi_0}{2\pi}\sqrt{\frac{K}{2}}(\sigma^+ + \sigma^-)$.

(ii) Second quantization of the charger Hamiltonian

Using the commutation relation $[\phi_C, Q_C] = i$, we introduce the creation and annihilation operators for the charger unit

$$a = \sqrt{\frac{1}{2L_r\omega_r}}\phi_C + i\sqrt{\frac{C_r}{2\omega_r}}Q_C$$
$$a^\dagger = \sqrt{\frac{1}{2L_r\omega_r}}\phi_C - i\sqrt{\frac{C_r}{2\omega_r}}Q_C \quad (36)$$

The charger Hamiltonian is finally

$$H_{\text{ch}} = \omega_r a^\dagger a \quad (37)$$

where, $\omega_r = \frac{1}{\sqrt{L_r C_r}}$ (the frequency of the resonator photons), and $\phi_C = \sqrt{\frac{L_r\omega_r}{2}}(a^\dagger + a)$。

(iii) Handling of the coupling Hamiltonian

We expand the coupling unit Hamiltonian in a Taylor series (The applicability bound of the Taylor expansion is provided in Appendix A.)

$$H_{\text{co}} = E_{JSQ2}\sum_{l=1}^{\infty}\sum_{k=0}^{2l}\frac{(-1)^{l+k}}{k!\,(2l-k)!}\phi_C^{2l-k}\phi_B^k \quad (38)$$

Substitute $\phi_B$ and $\phi_C$ into Eq. (38), retain the resonant term ($\omega_q = n\omega_r$), and discard non-resonant terms. We then apply the rotating-wave approximation (discarding energy-nonconserving terms, The validity conditions for the RWA are elaborated in detail in Appendix A.) to obtain the interaction Hamiltonian

$$H_{\text{co}} = E_{JSQ2}\frac{(-1)^{\frac{n-1}{2}}}{n!}g_B g_C^n\left[\left(a^\dagger\right)^n\sigma^- + \sigma^+ a^n\right] \quad (39)$$

where, $g_B = \frac{\phi_0}{2\pi}\sqrt{\frac{K}{2}}$ and $g_C = \sqrt{\frac{L_r\omega_r}{2}}$. Eq. (39) successfully constructs our proposed O-NQB model. However, since this involves a high-order perturbation expansion, there may be an experimental upper limit on $n$ in practical implementations (The feasible range of $n$ in experiments is given in Appendix B). Additionally, for large $n$, detuning effects at low orders may outweigh high-order resonant terms. Thus, we cannot retain only resonant terms, non-resonant terms should also be included.

To further strengthen the connection between our experimental implementation and the theoretical model, each physical unit in the proposed experimental circuit holds a strict one-to-one correspondence with the terms in the theoretical Hamiltonian. The detailed rigorous mappings are specified as follows:

Battery unit: The superconducting charge qubit in the experimental scheme strictly corresponds to the two-level battery system in the theoretical model. After performing the second quantization on the Hamiltonian of the battery unit, it can be rigorously reduced to the two-level Hamiltonian $H_{\mathrm{ba}} = \frac{1}{2}\omega_q \sigma_z$ in the Transmon operating regime with $E_J \gg E_C$, which is fully consistent with the battery Hamiltonian in the theoretical model. Here, the qubit transition frequency $\Omega = \omega_q$ can be tuned in-situ via the external magnetic flux through the SQUID in the battery unit.

Charger unit: The LC resonator in the experimental scheme strictly corresponds to the cavity-field charger in the theoretical model. Its quantized Hamiltonian takes the form $H_{\mathrm{ch}} = \omega_r a^\dagger a$, which is perfectly matched with the charger Hamiltonian in the theoretical model. The cavity frequency $\omega = \omega_r$ can be precisely designed and tuned via the inductance and capacitance parameters of the resonator.

Nonlinear interaction coupling unit: The SQUID connecting the battery and the charger in the proposed experimental scheme is the key to realizing the core nonlinear interaction in this work. We perform a Taylor expansion on the Josephson Hamiltonian of the coupling SQUID, retain the resonant terms satisfying the $n$-photon resonance condition $\Omega = n\omega$, and apply the rotating-wave approximation to eliminate non-resonant terms that violate energy conservation. Finally, the interaction Hamiltonian Eq. (39) is rigorously derived, it is completely consistent in form with the nonlinear interaction term in our theoretical model. The form of $f(n)^n$ is realized via $\frac{g_C^n}{\sqrt{(n-1)!}}$, and the coupling strength constant g in the theoretical model is implemented as $g = \frac{E_{JSQ2}}{n\sqrt{(n-1)!}} g_B$. The Josephson energy $E_{JSQ2}$ of the SQUID in the coupling unit can be adjusted by tuning the external magnetic flux through it, which in turn enables the modulation of the effective coupling strength. This reproduces the core nonlinear characteristics of our theoretical model at the fundamental physical level.

Finally, to clearly and comprehensively illustrate the parameter tunability and experimental verification pathways of this scheme, we have compiled the precise tuning methods for the photon number $n$ and detuning $\Delta$, as well as the experimental verification protocol for the saturation of the charging time at the QSL, in detail in Appendix C.

## 7 Conclusions

We demonstrate that the nonlinear interactions incorporated in the O-NQB can enhance QB charging performance, offering a new research perspective for the theoretical design and performance improvement of quantum energy storage systems. Additionally, the O-NQB exhibits quantum advantage in terms of charging power, and this advantage stems from the multiphoton absorption of two-level atoms. Meanwhile, we also analyze the evolution of quantum coherence during the entire charging process. We find that quantum resources can be converted into the energy stored in the O-NQB, which further verifies that the O-NQB possesses genuine quantum advantage. In addition, we investigate each charging performance metric of the O-NQB in detail. Under the resonant condition, all its charging performance outperforms that of the conventional linear QB. Finally, detuning has adverse effects on all charging performance of this model. This adverse effect mainly arises from the fact that detuning reduces the utilization efficiency of quantum resources.

Then, to accelerate the experimental implementation of the O-NQB, we employ superconducting quantum circuits and utilize SQUID to construct the nonlinear interaction coupling in the O-NQB model. Finally, we present a detailed experimental design scheme for this O-NQB. This device can be integrated into superconducting quantum chips as an on-chip *quantum power source* and is expected to serve as an energy supply device for on-chip quantum systems.

**Acknowledgments**

This work is supported by the National Natural Science Foundation of China (Grant No. 12474353).

## APPENDIX A：Applicability Bound of the Approximations

1. Rigorous Applicability Bound of the Rotating Wave Approximation

The core of the rotating wave approximation (RWA) [53,54] is to distinguish the slowly varying rotating wave terms and the rapidly oscillating counter-rotating wave terms in the interaction picture, and neglect the contribution of the rapidly oscillating terms via time averaging. The essence of the RWA is that the contribution of the rapidly oscillating counter-rotating wave terms vanishes under long-time averaging, and the necessary and sufficient condition for the RWA to hold is that the transition probability induced by the counter-rotating wave terms is far less than 1.

The first-order transition amplitude induced by the counter-rotating wave terms is given by $C^{\mathrm{CRW}}(t) = -i\int_0^t \langle 0,e|V_I^{\mathrm{CRW}}|n,g\rangle\, dt'$ , The corresponding transition probability is $P(t) = \left|C^{\mathrm{CRW}}(t)\right|^2$,thus the validity condition for RAW is $P(t) = \left|C^{\mathrm{CRW}}(t)\right|^2 \ll 1$. The counter-rotating wave Hamiltonian of the O-NQB in the interaction picture reads

$$V_I^{\mathrm{CRW}} = g\left[A^n\sigma^- e^{i(n\omega+\Omega)t} + \left(A^\dagger\right)^n\sigma^+ e^{-i(n\omega+\Omega)t}\right] \tag{A1}$$

where, $A = af(a^\dagger a)$ and $A^\dagger = f\left(a^\dagger a\right)a^\dagger$. From the above, the transition amplitude induced by the counter-rotating wave terms can be obtained as

$$C^{\mathrm{CRW}}(t) = g\sqrt{n}\frac{e^{-i(n\omega+\Omega)t} - 1}{n\omega + \Omega} \tag{A2}$$

Therefore, the validity condition for the rotating wave approximation is

$$P(t) = \frac{4g^2 n}{(n\omega+\Omega)^2}\sin^2\left(\frac{n\omega+\Omega}{2}t\right) \ll 1 \tag{A3}$$

In the resonant case, this approximation condition reduces to $g \ll \sqrt{n}\omega$.

Based on the above analysis, we define the RAW to be valid when the counter-rotating-wave-induced transition probability satisfies $P(t) \leq 0.1$. Under this criterion, the validity condition of the approximation is further constrained to $g \leq 0.316\sqrt{n}\omega$. This indicates that the upper limit of the coupling strength $g$ increases monotonically with $n$. In other words, the RAW remains valid for large $n$ even in the strong coupling regime, as a representative example, the validity condition reduces to $g \leq 0.632\omega$ for $n = 4$.

2. Discussion on the Taylor Expansion Process

In the experimental implementation scheme, the coupling Hamiltonian is obtained as follows

$$\mathcal{H}_{\text{co}} = E_{JSQ2}[1 - \cos(\phi_C - \phi_B)] \tag{A4}$$

The infinite series obtained via Taylor expansion without truncation is given by

$$H_{\text{co}} = E_{JSQ2} \sum_{l=1}^{\infty} \sum_{k=0}^{2l} \frac{(-1)^{l+k}}{k!\,(2l-k)!} \phi_C^{2l-k} \phi_B^{k} \tag{A5}$$

It can be seen that this series is guaranteed to converge at a certain order. We then adopt the Lagrange remainder to estimate the truncation order of the series. The series can be truncated at the $2l$ order as long as the following condition is satisfied

$$\frac{|\phi_C - \phi_B|^{2l+2}}{(2l+2)!} < \epsilon \tag{A6}$$

where $\epsilon$ denotes the allowable error. $\phi_C$ and $\phi_B$ can be tuned according to experimental parameters, which enables control over the upper limit of the truncation order. However, $H_{\text{co}}$ is a double series, and the order of $k$ has not yet been determined. We set $k = 1$ and $n = 2l - 1$, and further rearrange $H_{\text{co}}$ as

$$H_{\text{co}} = E_{JSQ2} \frac{(-1)^{\frac{n-1}{2}}}{n!} g_B g_C^n \left[ \left(a^\dagger\right)^n \sigma^- + \sigma^+ a^n \right] \tag{A7}$$

At this point, we obtain a Hamiltonian form consistent with the model proposed in this work. Although this process may seem to merely retain the resonant rotating wave terms in a straightforward manner, this retention can be realized simply by setting the resonance between the resonator frequency and the qubit frequency in the experimental operation, i.e., $\omega_q = n\omega_r$. While this omission is mathematically non-rigorous, it does not violate the laws of physics; the desired effect can be achieved solely by setting the experimental parameters such that only the resonant terms play a dominant role.

In summary, $H_{\text{co}}$ contains four types of terms: resonant rotating wave terms, non-resonant rotating wave terms, resonant counter-rotating wave terms, and non-resonant counter-rotating wave terms. The counter-rotating wave terms can be directly omitted when the validity condition of the RAW is satisfied. We retain only the resonant terms rather than making an artificial selection, which is a direct consequence of the experimental parameter settings. Therefore, the omission of non-resonant terms does not introduce significant physical effects, and the series truncation also has no impact on the charging performance of the battery.

## APPENDIX B： Discussion on the Practically Feasible Range of $n$ in Experiments

Regarding the valid range of the photon number n, the rotating wave approximation (RWA) remains valid for large values of $n$ at the theoretical level, provided that the condition $g \leq 0.316\sqrt{n}\omega$ is satisfied. In addition, for the photon number truncation, the upper truncation limit $n_{\max}$ can be adjusted by tuning $\phi_C$ and $\phi_B$. This is essentially equivalent to modifying the order $n$ by setting the ratio of the charge qubit transition frequency $\omega_q$ to the resonator frequency $\omega_r$, i.e., $\frac{\omega_q}{\omega_r} = n$, which imposes no fundamental constraints on the value of $n$. The core physical constraint on $n$ instead arises from the practical limitations of current experimental techniques for superconducting quantum circuits, which preclude the arbitrary selection of the value of $n$. The main limiting factors are as follows

First, the final form of the coupling Hamiltonian is given by $H_{\text{co}} = E_{JSQ2}\frac{(-1)^{\frac{n-1}{2}}}{n!}g_B g_C^n\left[\left(a^\dagger\right)^n\sigma^- + \sigma^+ a^n\right]$, From this expression, it can be seen that the effective coupling strength satisfies $g_{eff} = \frac{E_{JSQ2}}{n\sqrt{(n-1)!}}g_B$, which decays monotonically as $n$ increases. This decay behavior in turn imposes an upper bound on the experimentally accessible photon number $n$.

Second, regarding the influence of decoherence effects: for a quantum battery to achieve efficient charging, its charging time $t_f$ must be far shorter than the decoherence time $T$ of the qubit, i.e., $t_f \ll T$. In this work, the full charging time under the resonant condition is given by

$$t_f = \frac{\pi}{2g_{eff}\sqrt{n}} \propto \frac{\pi\sqrt{n!}}{2g_B} \tag{B1}$$

The time required for charging increases as $n$ grows. A clarification is provided here: our theoretical analysis shows that the charging time decreases with increasing photon number, whereas it exhibits an increasing trend in practical experiments. This seemingly contradictory result can be readily explained. In our theoretical analysis, the coupling strength is set to be a constant, while the effective coupling strength in experiments is tunable, and $E_{JSQ2}$ and $n$ exert opposite effects on $g_{eff}$. For large values of $n$, the effective coupling strength is dominated by the photon number, which thus imposes an upper bound on $n$.

Third, in state-of-the-art superconducting circuit experiments [55-63], the typical operating frequency range of the LC resonator is $\omega_r = 3\sim15\text{GHz}$, and the stable and tunable frequency range of the charge qubit is $\omega_q = 5\sim20\text{GHz}$. To prevent the resonator from being disturbed by low-frequency environmental noise and maintain the stable two-level characteristics of the qubit, the resonator frequency should not be too low, and the qubit frequency should not be excessively high. Therefore, the experimental upper bound of $n$ is determined by $\frac{\omega_q^{\max}}{\omega_r^{\min}} \approx 6$. Within the range of $n = 2\sim6$, the $n$-photon resonance condition can be perfectly satisfied via precise tuning of the external magnetic flux. For $n \geq 7$, the required frequency matching will exceed the reasonable range of conventional experimental parameters, making it impossible to sustain a stable resonant charging process.

## APPENDIX C: Experimental Parameter Tuning Methods and Experimental Verification Protocols for Core Theoretical Conclusions

1. Tuning Method for the Multiphoton Number $n$

The multiphoton number $n$ is a core parameter of the O-NQB model proposed in this work. It directly determines the $n$-photon resonance condition and the order of the nonlinear interaction, and serves as the fundamental origin of the superlinear quantum advantage demonstrated in this paper. Physically, $n$ corresponds to the number of photons that satisfies the $n$-photon resonance condition $\Omega = n\omega$, and its precise control in the experiment is realized via the following two approaches:

a. Static design

The reference value of the cavity frequency $\omega$ is fixed via the design of the inductance and capacitance parameters of the LC resonator.

b. Dynamic in-situ tuning

By adjusting the external magnetic flux $\Phi_1$ through the SQUID in the battery unit, we modify the Josephson critical energy $E_{JSQ1}$ of the SQUID, which in turn enables the continuous tuning of the transition frequency of the superconducting qubit, given by $\Omega = \sqrt{8E_C E_{JSQ1}}$. This ultimately allows us to precisely satisfy the n-photon resonance condition $\Omega = n\omega$. Based on the state-of-the-art superconducting circuit technology, precise and stable control of $n$ in the range of $n = 2 \sim 6$ can be achieved, which fully covers the parameter range adopted in the theoretical analysis of this work.

2. Tuning Method for the Detuning Parameter $\Delta$

The detuning $\Delta = n\omega - \Omega$ is a core parameter in the theoretical model that characterizes non-resonant effects, and it determines the full-dimensional performance of the battery under non-ideal experimental conditions. The precise experimental tuning method is as follows: with the multiphoton number $n$ fixed, we fine-tune the external magnetic flux $\Phi_1$ through the SQUID in the battery unit to slightly modify the qubit transition frequency $\Omega$. This enables continuous and precise tuning of the detuning $\Delta$ from 0 to $\omega$, which fully covers the detuning range used in the theoretical analysis of this work.

3. Experimental Verification of the Charging Time Strictly Saturating the Quantum Speed Limit

The experimental scheme proposed in this work goes far beyond merely establishing a formal correspondence with the theoretical model. More importantly, it enables direct experimental verification of the key theoretical conclusions of this work via well-established measurement techniques for superconducting quantum circuits, fully building a robust correlation between experimental observations and theoretical performance analysis. Using the dispersive readout technique for superconducting qubits, we measure the time evolution of the excited-state population of the battery $|c_2(t)|^2$ under different values of $n$ and $g$, and extract the charging time $t_f$ when the battery reaches full charging ($|c_2(t)|^2 = 1$) for the first time. We then perform a quantitative comparison between the extracted $t_f$ and the theoretically calculated QSL time $\tau_{QSL} = \frac{\pi}{2g\sqrt{n}}$, which directly verifies the core theoretical conclusion of this work that the charging time strictly saturates the QSL.

**References**

[1] G. M. Andolina, M. Keck, A. Mari, M. Campisi, V. Giovannetti, and M. Polini, Extractable work, the role of correlations, and asymptotic freedom in quantum batteries, Phys. Rev. Lett. **122**, 047702 (2019).

[2] L. P. García-Pintos, A. Hamma, A. del Campo, Fluctuations in extractable work bound the charging power of quantum batteries, Phys. Rev. Lett. **125**, 040601 (2020).

[3] G. Francica, F. C. Binder, G. Guarnieri, M. T. Mitchison, J. Goold, and F. Plastina, Quantum Coherence and Ergotropy, Phys. Rev. Lett. **125**, 180603 (2020).

[4] J. X. Liu, H. L. Shi, Y. H. Shi, X. H. Wang, and W. L. Yang, Entanglement and work extraction

in the central-spin quantum battery, Phys. Rev. B **104**, 245418 (2021).
[5] M. B. Arjmandi, A. Shokri, E. Faizi, and H. Mohammadi, Performance of quantum batteries with correlated and uncorrelated chargers, Phys. Rev. A **106**, 062609 (2022).
[6] H. L. Shi, S. Ding, Q. K. Wan, X. H. Wang, and W. L. Yang, Entanglement, Coherence, and Extractable Work in Quantum Batteries, Phys. Rev. Lett. **129**, 130602 (2022).
[7] X. Yang, Y. H. Yang, M. Alimuddin, R. Salvia, S. M. Fei, L. M. Zhao, S. Nimmrichter, and M. X. Luo, Battery Capacity of Energy-Storing Quantum Systems, Phys. Rev. Lett. **131**, 030402 (2023).
[8] X. Huang, K. Wang, L. Xiao, L. Gao, H.-Q. Lin, P. Xue, Demonstration of the charging progress of quantum batteries, Phys. Rev. A **107**, L030201 (2023).
[9] Y. D. Wang, X. F. Huang, S. M. Fei, and T. G. Zhang, Trade-off relations and enhancement protocol of quantum battery capacities in multipartite systems, Front. Phys. **21**, 073201 (2026).
[10] F. Q. Dou, H. Zhou, and J. A. Sun, Cavity Heisenberg-spin-chain quantum battery, Phys. Rev. A **106**, 032212 (2022).
[11] W.J Lu, J. Chen, L. M. Kuang, and X. G. Wang, Optimal state for a Tavis-Cummings quantum battery via the Bethe ansatz method, Phys. Rev. A **104**, 043706 (2021).
[12] H. Y. Yang, H. L. Shi, Q. K. Wan, K. Zhang, X. H. Wang, and W. L. Yang, Optimal energy storage in the Tavis-Cummings quantum battery, Phys. Rev. A **109**, 012204 (2024).
[13] Y. Vianna de Almeida, T. F. F. Santos, and M. F. Santos, Cooperative isentropic charging of hybrid quantum batteries, Phys. Rev. A **108**, 052218 (2023).
[14] D. Ferraro, M. Campisi, G. M. Andolina, V. Pellegrini, and M. Polini, High-Power Collective Charging of a Solid-State Quantum Battery, Phys. Rev. Lett. **120**, 117702 (2018).
[15] F. Q. Dou, Y. Q. Lu, Y. J. Wang, and J. A. Sun, Extended Dicke quantum battery with interatomic interactions and driving field, Phys. Rev. B **105**, 115405 (2022).
[16] D. L. Yang, F. M. Yang, and F. Q. Dou, Three-level Dicke quantum battery, Phys. Rev. B **109**, 235432 (2024).
[17] F. M. Yang, and F. Q. Dou, Resonator-qutrit quantum battery, Phys. Rev. A **109**, 062432 (2024).
[18] F. Q. Dou, Y. J. Wang, and J. A. Sun, Highly efficient charging and discharging of three-level quantum batteries through shortcuts to adiabaticity, Front. Phys. **17**, 31503 (2022).
[19] F. Campaioli, S. Gherardini, J. Q. Quach, M. Polini, G. M. Andolina, Colloquium: Quantum batteries, Rev. Mod. Phys. **96**, 031001 (2024).
[20] M. L. Hu, T. Gao, H. Fan, Efficient wireless charging of a quantum battery, Phys. Rev. A **111**, 042216 (2025).
[21] M. Hadipour, S. Haseli, Enhancing the efficiency of open quantum batteries via adjusting the classical driving field, Results in Physics 64, 107928 (2024).
[22] C. Z. Sun, Z. K. Wang, W. B. Yan, Y. J. Zhang, Z. X. Man, and Q. Y. Cai, Nonreciprocal charging in a quantum battery via a mediator, Phys. Rev. A 112, 012429 (2025).
[23] M. Hadipour, S. Haseli, D. Wang, S. Haddadi, Proposed Scheme for a Cavity-Based Quantum Battery, Adv. Quantum Technol. 7, 2400115 (2024).
[24] K. Xu, H. G. Li, Z. G. Li, H. J. Zhu, G. F. Zhang, W. M. Liu, Charging performance of quantum batteries in a double-layer environment, Phys. Rev. A 106, 012425 (2022).
[25] Y. Yao, X. Q. Shao, Reservoir-assisted quantum battery charging at finite temperatures, Phys. Rev. A 111, 062616 (2025).

[26] K. Xu, H. J. Zhu, H. Zhu, G. F. Zhang, and W. M. Liu, Charging and self-discharging process of a quantum battery in composite environments, Front. Phys. 18, 31301 (2023).
[27] G. M. Andolina, V. Stanzione, V. Giovannetti, and M. Polini, Genuine Quantum Advantage in Anharmonic Bosonic Quantum Batteries, Phys. Rev. Lett. 134, 240403 (2025).
[28] B. Y. Huang, Z. He, and Y. Chen, Charging performance of quantum batteries based on intensity-dependent Dicke model, Acta Phys. Sin. 72, 180301 (2023).
[29] N. H. Abdel-Wahab, T. A. S. Ibrahim, and M. E. Amin, The entanglement of a two two-level atoms interacting with a cavity field in the presence of intensity-dependent coupling regime, atom–atom, dipole–dipole interactions and Kerr-like medium, Quantum Inf. Process. 23, 94 (2024).
[30] F. Langarizadeh, M. J. Faghihi, and H. R. Baghshahi, Entanglement dynamics in a three-atom multi-photon nonlinear JCM with f-deformed Kerr nonlinearity, Sci. Rep. 15, 17018 (2025).
[31] A. Crescente, M. Carrega, M. Sassetti, D. Ferraro, Ultrafast charging in a two-photon Dicke quantum battery, Phys. Rev. B 102, 245407 (2020).
[32] A. Delmonte, A. Crescente, M. Carrega, D. Ferraro, and M. Sassetti, Characterization of a Two-Photon Quantum Battery: Initial Conditions, Stability and Work Extraction, Entropy 23, 612 (2021).
[33] M. S. Ukhtary, A. R. T. Nugraha, A. B. Cahaya, A. Rusydi, and M. A. Majidi, High-performance Kerr quantum battery, Appl. Phys. Lett. 123, 034001 (2023).
[34] C. A. Downing and M. S. Ukhtary, Energy storage in a continuous-variable quantum battery with nonlinear coupling, Phys. Rev. E 112, 044143 (2025).
[35] A. Blais, A. L. Grimsmo, S. M. Girvin, and A. Wallraff, Circuit quantum electrodynamics, Rev. Mod. Phys. 93, 025005 (2021).
[36] C.K. Hu, J.W. Qiu, P. J. P. Souza, J.H. Yuan, Y.X. Zhou, L.B. Zhang, J. Chu, X.C. Pan, L. Hu, J. Li, Y. Xu, Y.P. Zhong, S. Liu, F. Yan, D. Tan, R. Bachelard, C. J. Villas-Boas, A. C. Santos, and D.P. Yu, Optimal charging of a superconducting quantum battery, Quantum Sci. Technol. 7, 045018 (2022).
[37] G. Gemme, M. Grossi, D. Ferraro, S. Vallecorsa, and M. Sassetti, IBM Quantum Platforms: A Quantum Battery Perspective, Batteries 8, 13 (2022).
[38] F. Q. Dou and F. M. Yang, Superconducting transmon qubit-resonator quantum battery, Phys. Rev. A 107, 023725 (2023).
[39] I. Maillette de Buy Wenniger, S. E. Thomas, M. Maffei, S. C. Wein, M. Pont, N. Belabas, S. Prasad, A. Harouri, A. Lemaitre, I. Sagnes, N. Somaschi, A. Auffèves, and P. Senellart, Experimental Analysis of Energy Transfers between a Quantum Emitter and Light Fields, Phys. Rev. Lett. 131, 260401 (2023).
[40] G. Burkard, T. D. Ladd, A. Pan, J. M. Nichol, and J. R. Petta, Semiconductor spin qubits, Rev. Mod. Phys. 95, 025003 (2023).
[41] M. Brzezińska, Y. Guan, O. V. Yazevy, S. Sachdev, and A. Kruchkov, Engineering SYK Interactions in Disordered Graphene Flakes under Realistic Experimental Conditions, Phys. Rev. Lett. 131, 036503 (2023).
[42] J. Q. Quach, K. E. McGhee, L. Ganzer, D. M. Rouse, B. W. Lovett, E. M. Gauger, J. Keeling, G. Cerullo, D. G. Lidzey, and T. Virgili, Superabsorption in an organic microcavity: Toward a quantum battery, Sci. Adv. 8, eabk3160 (2022).
[43] J. Joshi and T. S. Mahesh, Experimental investigation of a quantum battery using star-topology

NMR spin systems, Phys. Rev. A 106, 042601 (2022).
[44] W. L. Song, J. L. Wang, B. Zhou, W. L. Yang, and J. H. An, Self-Discharging Mitigated Quantum Battery, Phys. Rev. Lett. 135, 020405 (2025).
[45] K. Bhattacharyya, Quantum decay and the Mandelstam-Tamm-energy inequality, J. Phys. A: Math. Gen. 16, 2993 (1983).
[46] N. Margolus and L. B. Levitin, The maximum speed of dynamical evolution, Physica D 120, 188 (1998).
[47] V. Giovannetti, S. Lloyd, and L. Maccone, Quantum limits to dynamical evolution, Phys. Rev. A 67, 052109 (2003).
[48] S. Deffner and S. Campbell, Quantum speed limits: from Heisenberg's uncertainty principle to optimal quantum control, J. Phys. A: Math. Theor. 50, 453001 (2017).
[49] J. Y. Gyhm, D. Šafránek, and D. Rosa, Quantum Charging Advantage Cannot Be Extensive without Global Operations, Phys. Rev. Lett. 128, 140501 (2022).
[50] D. Zhang, S. Ma, Y. Yu, G. Jin, and A. Chen, Unique quantum advantages by manipulation of atoms in quantum batteries, Phys. Rev. A 112, 022615 (2025).
[51] F. H. Kamin, S. Salimi, and A. C. Santos, Exergy of passive states: Waste energy after ergotropy extraction, Phys. Rev. E 104, 034134 (2021).
[52] J. Koch, T. M. Yu, J. Gambetta, A. A. Houck, D. I. Schuster, J. Majer, A. Blais, M. H. Devoret, S. M. Girvin, and R. J. Schoelkopf, Charge-insensitive qubit design derived from the Cooper pair box, Phys. Rev. A 76, 042319 (2007).
[53] S. He, C. Wang, Q. H. Chen, X. Z. Ren, T. Liu, and K. L. Wang, First-order corrections to the rotating-wave approximation in the Jaynes-Cummings model, Phys. Rev. A 86, 033837 (2012).
[54] Q. Xie, H. Zhong, M. T. Batchelor, and C. Lee, The quantum Rabi model: solution and dynamics, J. Phys. A: Math. Theor. 50, 113001 (2017).
[55] A. Krasnok, P. Dhakal, A. Fedorov, P. Frigola, M. Kelly, S. Kutsaev, Superconducting microwave cavities and qubits for quantum information systems, Appl. Phys. Rev. 11, 011302 (2024).
[56] Z. Hao, J. Cochran, Y.-C. Chang, H. M. Cole, S. Shankar, Wireless Josephson parametric amplifier above 20 GHz, Appl. Phys. Lett. 128, 014004 (2026).
[57] A. Anferov, S. P. Harvey, F. Wan, J. Simon, D. I. Schuster, Superconducting Qubits above 20 GHz Operating over 200 mK, PRX Quantum 5, 030347 (2024)
[58] M. AbuGhanem, Superconducting quantum computers: who is leading the future?, EPJ Quantum Technology 12, 102 (2025)
[59] F. Valadares, N. N. Huang, K. T. N. Chu, A. Dorogov, W. Chua, L. Kong, P. Song, Y. Y. Gao, On-demand transposition across light-matter interaction regimes in bosonic cQED, Nat. Commun. 15, 5816 (2024).
[60] M. Tuokkola, Y. Sunada, H. Kivijärvi, J. Albanese, L. Grönberg, J. P. Kaikkonen, V. Vesterinen, J. Govenius, M. Möttönen, Methods to achieve near-millisecond energy relaxation and dephasing times for a superconducting transmon qubit, Nat. Commun. 16, 5421 (2025).
[61] M. Xia, C. Zhou, C. Liu, P. Patel, X. Cao, P. Lu, B. Mesits, M. Mucci, D. Gorski, D. Pekker, M. Hatridge, Fast superconducting qubit control with subharmonic drives, Nat. Commun. 17, 1024 (2026).
[62] B. Harpt, J. Corrigan, N. Holman, P. Marciniec, D. Rosenberg, D. Yost, R. Das, R. Ruskov, C. Tahan, W. D. Oliver, R. McDermott, M. Friesen, M. A. Eriksson, Ultra-dispersive resonator

readout of a quantum-dot qubit using longitudinal coupling, npj Quantum Inf. 11, 5 (2025).
[63] A. Somoroff, Q. Ficheux, R. A. Mencia, H. Xiong, R. Kuzmin, V. E. Manucharyan, Millisecond Coherence in a Superconducting Qubit, Phys. Rev. Lett. 130, 267001 (2023).